\documentclass{article}

\usepackage{arxiv}

\usepackage[utf8]{inputenc} 
\usepackage[T1]{fontenc}    
\usepackage{hyperref}       
\usepackage{url}            
\usepackage{booktabs}       
\usepackage{amsfonts}       
\usepackage{nicefrac}       
\usepackage{microtype}      
\usepackage{lipsum}		
\usepackage{graphicx}
\usepackage{natbib}
\usepackage{doi}
\usepackage{graphicx}
\usepackage{amsmath}
\usepackage{bm}
\usepackage[utf8]{inputenc} 
\usepackage[T1]{fontenc}    
\usepackage{hyperref}       
\usepackage{url}            
\usepackage{booktabs}       
\usepackage{txfonts}
\usepackage{algorithm}
\usepackage[noend]{algpseudocode}

\usepackage{amsfonts}       
\usepackage{nicefrac}       
\usepackage{microtype}      
\usepackage{lipsum}		
\usepackage{graphicx}
\usepackage{natbib}
\usepackage{doi}
\usepackage{mathtools}
\usepackage{todonotes}
\usepackage{tablefootnote}
\usepackage{caption}
\usepackage{subcaption}
\usepackage[inline]{enumitem}
\usepackage{tabu}
\usepackage{xcolor}
\hypersetup{
	colorlinks,
	linkcolor={red!50!black},
	citecolor={blue!50!black},
	urlcolor={blue!80!black}
}

\makeatletter
\def\cl@chapter{\@elt {theorem}}
\makeatother
\usepackage{cleveref}
\usepackage{dblfloatfix} 
\usepackage{tikz}

\def\checkmark{\tikz\fill[scale=0.4](0,.35) -- (.25,0) -- (1,.7) -- (.25,.15) -- cycle;} 

\newcommand*{\affaddr}[1]{#1} 
\newcommand*{\affmark}[1][*]{\textsuperscript{#1}}

\newcolumntype{C}[1]{>{\centering\arraybackslash}p{#1}}
\newcommand*{\email}[1]{\texttt{#1}}
\setcitestyle{numbers}

\title{FRC-TOuNN: Topology Optimization of Continuous Fiber Reinforced Composites using Neural Network} 

\author{%
	Aaditya Chandrasekhar \affmark[1],Amir Mirzendehdel \affmark[2], Morad Behandish \affmark[2], and Krishnan Suresh \affmark[1]\\
	\affaddr{\affmark[1]Department of Mechanical Engineering, University of Wisconsin - Madison}\\
	\affaddr{\affmark[2] Palo Alto Research Center, California }\\
	\email{\{achandrasek3, ksuresh\}@wisc.edu}\\
	\email{\{amir.mirzendehdel, moradbeh\}@parc.com}%
}

\date{}

\renewcommand{\shorttitle}{\textit{arXiv} Template}

\begin{document}
\maketitle

\begin{abstract}
    In this paper, we present a topology optimization (TO) framework to simultaneously optimize the matrix topology and fiber distribution of functionally graded continuous fiber-reinforced composites (FRC). Current  approaches in density-based TO for FRC use the underlying finite element mesh both for analysis and design representation. This poses several limitations while enforcing sub-element fiber spacing and generating high-resolution continuous fibers. In contrast, we propose a mesh-independent representation based on a neural network (NN) both to capture the matrix topology and fiber distribution. The implicit NN-based representation enables geometric and material queries at a higher resolution than a mesh discretization. This leads to the accurate extraction of functionally-graded continuous fibers. Further, by integrating the finite element simulations into the NN computational framework, we can leverage automatic differentiation for end-to-end automated sensitivity analysis, i.e., we no longer need to manually derive cumbersome sensitivity expressions. We demonstrate the effectiveness and computational efficiency of the proposed method through several numerical examples involving various objective functions. We also show that the optimized continuous fiber reinforced composites can be directly fabricated at high resolution using additive manufacturing.
\end{abstract}


\section{Introduction}
\label{sec:intro}

Fiber reinforced composites (FRCs) are comprised of a base matrix (example: polymers) with fiber inclusions (example: glass) \cite{gandhi2020discontinuous}. They exhibit a high stiffness-to-weight ratio and are deployed extensively in the automotive, robotics, medical, aeronautical, and aerospace industries \cite{DOMM_application_FRC}. With the advent of additive manufacturing (AM) \cite{Zhang2017}, applications of FRCs have  increased significantly \cite{liu2018current, parandoush2017review}. 

\begin{figure}[b]
	\begin{center}
		\includegraphics[width=0.5\linewidth]{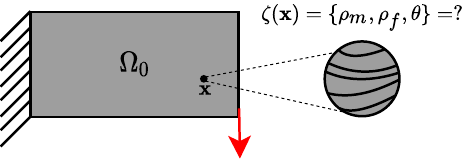}%
		\caption{FRC optimization involves 3 design variables: matrix density  ($\rho_m$), fiber density ($\rho_f$), and fiber orientation  ($\theta$) at each point.}
		\label{fig:domain_topOpt_problem}
	\end{center}
\end{figure}

FRCs can be designed using long  (i.e., continuous) fibers or short fibers; the former is  preferred in mission-critical applications due to their superior material properties, dimensional stability, and robustness in fiber orientation and bonding  \cite{thomason2002influence,hine2002numerical}. However, for long-fiber FRCs to be effective, the base matrix topology, the fiber-density (spacing between individual fibers), and the fiber-orientation must be optimized simultaneously (\Cref{fig:domain_topOpt_problem}) \cite{pedersen1989optimal, pedersen1990bounds, pedersen1991thickness}.

\begin{figure*}[]
	\begin{center}
		\includegraphics[scale=0.9,trim={0 0 0 0},clip]{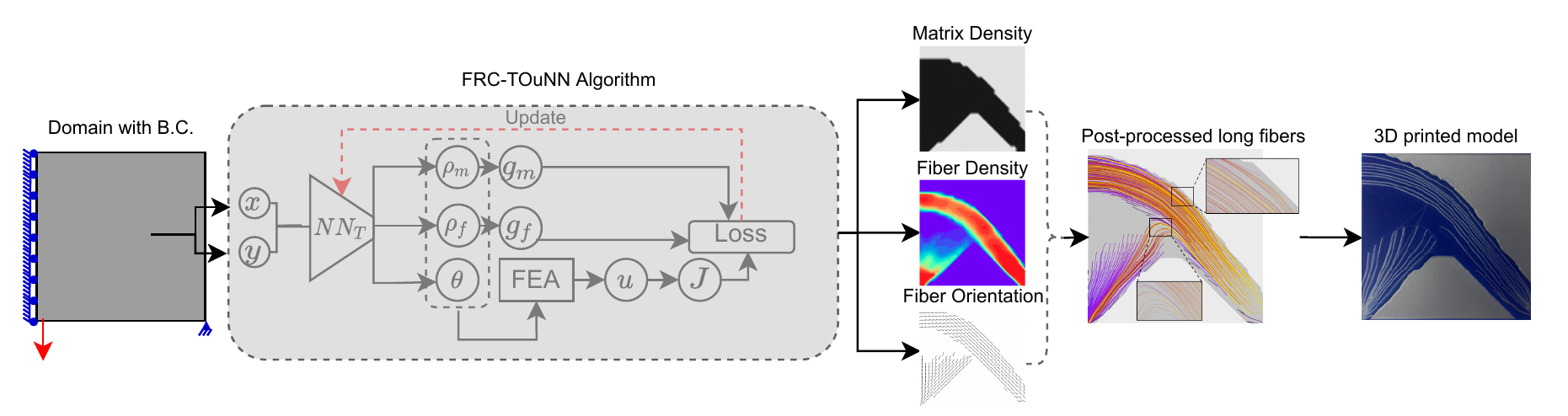}%
		\caption{Given the initial design domain and loading condition, we propose an algorithm that utilizes a mesh-independent neural-network-based representation to simultaneously optimize the matrix-topology, fiber-density, and fiber-orientation. This NN-based representation enables an end-to-end automatic differentiation, where we only need to define a loss function without manually deriving the sensitivity expressions. Further, this implicit representation allows realization of continuous fibers at high-resolution in the post-processing stage using the trained NN. Subsequently, the optimized FRC part can be fabricated using a 3D printer. }
		\label{fig:method_graphicalAbstract}
	\end{center}
\end{figure*}

Topology optimization (TO) is a computational design tool widely used during the early stages of design to automatically generate high-performance lightweight parts.
Various TO methods have been proposed for optimizing FRCs, based on different design parameters and formulations \cite{saketh20213PhaseFiberTO,lee2018topology,li2021full}. The most popular approach is based on Solid Isotropic Material with Penalization (SIMP) \cite{sigmund1997design}, where a fictitious pseudo-density field is defined over the underlying discretization is used for design representation, analysis, and optimization. Although in most practical cases, involving isotropic materials, we want every element to be either solid or void to ensure manufacturability, the pseudo-density at an element is allowed to take intermediate values ($0\le \rho_e \le 1$) during the optimization to ensure differentiability. It is well-established that an intermediate density value at a given point in the design ( i.e., macro-scale) exhibits the \emph{homogenized} properties of a microstructure at a lower scale \cite{Bendsoe2003}. For instance, the effective elastic modulus $\hat{E}$ for an element with intermediate density $\rho_e$ and penalization factor $p$ is approximated as $\hat{E} = \rho_e^p E_0$, where $E_0$ is the elastic modulus of the base material. In other words, instead of modeling the geometric features at a lower scale with elastic modulus $E_0$, we use the homogenized value $\hat{E}$ at the macro-scale to simplify the formulation and improve computational efficiency.

On the other hand, FRC structures are inherently multi-scale and the mesh-dependent representation of the matrix-topology and fiber layout can pose several limitations in the physical realization and manufacturing of the optimized part. To maintain computational efficiency, analysis and optimization are performed using homogenized properties defined by the ``rule of mixture''. For instance, given the volume fraction of fibers $\rho_f$ and elastic moduli for fibers $E_f$ and matrix $E_m$, we have:
\begin{equation}
	\hat{E} = \rho_f E_f + (1-\rho_f) E_m\quad 0\le \rho_f\le 1
\end{equation}

To realize the fiber inclusions at a lower scale, the design representation must be able to accommodate sub-element geometric and material queries. However, a discretized representation of FRC structures, where the fiber orientations are associated with each finite element, is not suitable for extracting high-resolution continuous fibers while enforcing sub-element fiber spacing --- to quote \cite{saketh20213PhaseFiberTO} \emph{ ``\dots element-wise fiber orientation is not amenable for manufacturing or 3D printing \dots''}.

In this paper, we rely on  a \emph{ mesh-independent neural-network (NN) representation} of the matrix-topology, fiber-density, and fiber-orientation. 
The implicit NN-based representation enables geometric and material queries at a higher resolution than the background mesh, which leads to accurate extraction of functionally-graded continuous fibers. Further, by integrating the physical simulations into the NN computational framework, we leverage automatic differentiation (AD) to provide end-to-end automated sensitivity analysis by extending the method presented in \cite{Chandrasekhar2021AuTO,ChandrasekharMMTOuNN2020} to continuous fiber reinforced composites. In other words, we eliminate the need to manually derive the sensitivity expressions and only need to define a loss function capturing the objective and constraints. This permits rapid exploration of material models and design objectives, especially for applications involving dynamic and nonlinear physics.

\Cref{fig:method_graphicalAbstract} provides an overview of the proposed method, where the spatial coordinates from the Euclidean space are mapped to the frequency space. Thus, the design is implicitly represented by a fully-connected feedforward NN with its output layer comprising three neurons corresponding to the matrix-topology, fiber-density, and fiber-orientation. As the finite element analysis (FEA) is performed within an NN framework (e.g., PyTorch), we leverage AD to compute the sensitivities and update the NN weights (e.g., Adam optimizer). Once the NN is \emph{trained}, i.e., once the optimization is complete, a high-resolution FRC can be extracted and fabricated, for instance, using additive manufacturing (AM).

\section{Literature Review}
\label{sec:litReview}

\subsection{FRC Optimization Methods}
\label{sec:litRev}

While analytical methods have been developed for optimizing FRCs \cite{bendsoe1994analytical,cheng1994improved,luo1998optimal}, the most popular method for optimizing FRCs is through topology optimization (TO). Current TO methods include density based   \cite{Bendsoe2003,Bendsoe2008,sigmund2001Code99,sigmund2013topology}, level-set \cite{Wang2003LevelSet}, topological sensitivity  \cite{Suresh2013LevelSet,Mirzendehdel2018Anisotropy,mirzendehdel2016support}, and evolutionary methods \cite{rozvany2009criticalReviewTO}. Among these, the density based methods are the most popular.

Optimizing the matrix-topology using TO is fairly straightforward and is not discussed here~\cite{Bendsoe2008}. Instead, we consider TO methods for optimizing the fiber-density  and fiber-orientation.  These methods can be  classified as (a) continuous fiber angle optimization (CFAO), (b) discrete material optimization (DMO), or  (c) free material optimization (FMO). 
In CFAO, the orientation of the fiber in each element is treated as a continuous design variable \cite{lindgaard2011optimization,xia2017optimization}.  For instance the simultaneous optimization of AM build orientation, topology and fiber orientation was reported \cite{Chandrasekhar2020Fiber}, while the simultaneous optimization of the topology and oriented material microstructure was reported in \cite{yan2019concurrent}.  A common characteristics of these methods was that  matrix at an element was either void, or present with oriented fibers. Recently,  \cite{li2021full} proposed a full-scale method for simultaneous design of matrix topology and fiber orientation; however, the spacing was fixed. In contrast to these two-phase approaches (i.e., void and anisotropic material), a three-phase approach was proposed in \cite{lee2018topology} wherein fiber concentration was considered along with material fraction as design variables. A topological derivative based method was proposed in \cite{saketh20213PhaseFiberTO} with binary-valued fiber concentration at each point.
In contrast to CFAO, DMO methods \cite{stegmann2005discrete}  restrict the orientation to predetermined discrete values. While DMO approaches are often simpler with better convergence, they may result in discontinuous fibers.  Finally, in FMO, the individual components of the elasticity tensor are treated as design variables, leading to theoretically optimal designs, but may lack a physical interpretation \cite{li2021full}. 

\subsection{FRC Post-Processing Methods}
Once an optimal FRC is computed, a post-processing step is required to determine the fiber paths or layout. A commonly used approach is the streamline methodology where fiber paths are obtained from the streamlines of the optimized fiber angle field \cite{khani2015optimum}. However, this does not guarantee uniform fiber distribution, and thus could lead to sub-optimal overlaps or gaps in the structure. Another common approach used in level set methods is to use  parallel offsets of the contours to generate the fibers \cite{Sivapuram2016}. This, however, reduces the design freedom by requiring the fibers to orient along the contours of the structure and may lead to sub-optimal results. Recently, continuous fiber generation based on the stripe patterns algorithm proposed by \cite{knoppel2015stripe} was used in \cite{boddeti2020optimal} that allowed for the generation of continuous fibers, considering both orientation and concentration, without optimizing the distribution of the underlying matrix.

\subsection{Application of Neural-Network in TO}
Since the proposed method relies on NN for optimization, we review relevant literature. Applications of NN in TO can be  classified under two categories: (a) accelerating TO through NN (b) directly using NN to compute topologies. 

Under the former category, \cite{zhang2021speeding} proposed accelerating TO by predicting gradients via online training of a NN; \cite{Banga2018CNN_TO} proposed training a convolutional NN with a large database of optimized topologies and their corresponding loads to speed up topology prediction under novel loading scenario. Similar strategies were proposed in \cite{Sosnovik2019TONN} and \cite{ulu2016dataTO}. \cite{Nie2020TO_GAN} proposed using a generative adversarial network to train a generator to predict topologies conditioned under different loading scenarios. \cite{Lin2018} employed NN to recognize and replace evolving features, thereby reducing number of TO iterations. \cite{white2019multiscale} and \cite{wang2020deep} trained autoencoders with data pertinent to homogenized microstructures. The encoders were then employed in a TO framework, thereby eliminating the need to perform homogenization during the optimization loop. 

In contrast to training NNs with data,  one can use them  directly to capture the topology as implicit function of the spatial coordinates, by recognizing that NNs are nothing but functions capable of capturing complex signals. One approach is based on continuous coordinate networks whose representational capacity is limited by the network rather than the grid size. They have been recently employed for physics simulations \cite{Raissi2019}, modeling 3D shapes \cite{gropp2020implicit}, view synthesis \cite{mildenhall2020nerf}, semantic labeling \cite{kohli2020semantic}, texture synthesis \cite{oechsle2019texture} etc. Recently, coordinate NNs have been used for isotropic TO \cite{ChandrasekharTOuNN2020}, \cite{hoyer2019neural}. The current work  extends this work for the optimization of FRCs. 

\subsection{Contributions and Outline}
\label{sec:intro_contributions}
The proposed method can be classified as a CFAO method based on a neural-network representation to capture the matrix-topology, the  fiber-orientation, and fiber-density. 
We summarize and contrast the capabilities of the proposed method against previous work in \Cref{table:recentLitComparison}.

\begin{table}[]
	\caption{Comparison with recent literature}
	\begin{center}
		\begin{tabular}{  l | C{12mm} C{12mm} C{12mm}  }
			\hline \hline
			
			Reference & Matrix Topology & Fiber Density & Fiber Orientation   \\ \hline
			\cite{saketh20213PhaseFiberTO}, \cite{Chandrasekhar2020Fiber}, \cite{Stegmann2005}, \cite{brampton201} & $\checkmark$ & $\times $ & $\checkmark $  \\ \hline
			
			\cite{lee2018topology}, \cite{papapetrou2020stiffness}& $\checkmark$ & $\checkmark $ & $\times $  \\ \hline
			
			\cite{li2021full}, \cite{Steuben2016CADImplicitSlicing}, \cite{pedersen1991thickness} & $\times$ & $\times $ & $\checkmark $  \\ \hline

			Proposed  & $\checkmark $ & $\checkmark $ & $\checkmark$ \\ \hline
		\end{tabular}
	\end{center}
	\label{table:recentLitComparison}
\end{table}

The main contributions of this paper are: 
\begin{itemize}
	\item Introducing an implicit NN-based representation for FRCs that enables realization of continuous fibers at a high resolution.
	\item Expressing all computations including FEA in an end-to-end automatic differentiable framework  for automated sensitivity analysis.
	\item Proposing a density-based TO approach by simultaneously optimizing all three design variables, while also reducing the number of design variables.
	\item Presenting a new post-processing method for extraction of continuous long fibers at a high resolution.
\end{itemize}

In Section \ref{sec:method}, we describe the proposed optimization method and subsequent post-processing stage for continuous fiber extraction.
In Section \ref{sec:numExpts} we demonstrate the validity, robustness, and efficiency of the framework. Finally, research challenges, opportunities and conclusions are summarized in Section \ref{sec:conclusion}.

\section{Literature Review}
\label{sec:litRev}

\subsection{Topology Optimization of FRCs }
\label{sec:litRev_FRCTO}
As mentioned earlier, While analytical methods have been developed for optimizing FRCs \cite{bendsoe1994analytical,cheng1994improved,luo1998optimal}, the most popular method for optimizing FRCs is through TO. Current TO methods include density-based   \cite{Bendsoe2003}, \cite{Bendsoe2008}, \cite{sigmund2001Code99, sigmund2013topology}, level-set \cite{Wang2003LevelSet}, topological sensitivity  \cite{Suresh2013LevelSet}, \cite{Mirzendehdel2018Anisotropy} and evolutionary methods \cite{rozvany2009criticalReviewTO}. Among these, the density-based methods are the most popular.

Optimizing the matrix topology using TO is fairly straightforward and is not discussed here\cite{Bendsoe2008}. Instead, we consider TO methods for optimizing the fiber-density  and fiber-orientation.  These methods can be  classified as (a) continuous fiber angle optimization (CFAO), (b) discrete material optimization (DMO), and (c) free material optimization (FMO).

In CFAO, the orientation of the fiber in each element is treated as a continuous design variable \cite{lindgaard2011optimization}, \cite{xia2017optimization}. For instance, the simultaneous optimization of AM build orientation, topology and fiber orientation was reported \cite{Chandrasekhar2020Fiber}. The simultaneous optimization of the topology and oriented material microstructure was reported in \cite{yan2019concurrent}.  A key factor was the models considered the matrix to be fully embedded with fibers. In other words, the topology either consisted of matrix fully embedded with oriented fibers or void. Recently,  \cite{li2021full} proposed a full-scale method capable of simultaneous design for the topology, fiber path, and its morphology. However, the spacing was fixed throughout the design. In contrast to these two-phase approaches (i.e., void and anisotropic material), a three-phase approach was proposed in \cite{lee2018topology} wherein fiber concentration was considered along with material fraction in a given design space. A topological derivative based method was proposed in \cite{saketh20213PhaseFiberTO} with binary-valued fiber concentration at each point.

In contrast, DMO \cite{stegmann2005discrete}  restrict the orientation to predetermined discrete values. While DMO approaches are often simpler with better convergence, they may result in discontinuous fibers.  However, a DMO model was used which limits the practicality of the design as discussed. 

Finally, in FMO, the individual components of the elasticity tensor are treated as design variables, leading to theoretically optimal designs, but may lack a physical interpretation \cite{li2021full}.

Once an optimal topology is computed with fiber orientations, a post-processing step is utilized to determine the fiber paths or layout. A commonly used approach is the streamline methodology where fiber paths are obtained from the streamlines of the optimized fiber angle field \cite{khani2015optimum}. However, this does not guarantee uniform fiber distribution and thus could lead to sub-optimal overlaps or gaps in the compiled structure. Another common approach with level set based topology optimization is to use the parallel offsets of the contours of the optimized structure to orient the fibers \cite{Sivapuram2016}. This, however, reduces the design freedom by requiring the fibers to orient along the contours of the structure and may lead to sub-optimal results. Recently, continuous fiber compilation based on the stripe patterns algorithm proposed by \cite{knoppel2015stripe} was used in \cite{boddeti2020optimal} that allowed for the compilation of continuous fibers with consideration to orientation and concentration.

\subsection{ Neural-Network based TO }
\label{sec:NN_TO}

Since this paper exploits neural-networks (NNs) to carry out TO, we briefly review prior work in this discipline.  Applications of NN in TO can be  classified under three categories: (a) accelerating TO through NN,  (b) TO via NN-generated input-topology maps, and (c) directly using NN to compute topologies. The first category includes the work of \cite{Banga2018CNN_TO}, \cite{Sosnovik2019TONN}, \cite{ulu2016dataTO}. 
Proposed methods under the second category include the works of \cite{Nie2020TO_GAN}, \cite{Yu2019}, \cite{Lin2018}, \cite{elingaard2021homogenization}. Here, the generated topologies aren't guaranteed to be valid. The third approach rely on the fact that NNs are but implicit functions capable of capturing complex topologies \cite{hoyer2019neural, ChandrasekharTOuNN2020}.  

\subsection{Paper Contributions}
\label{sec:intro_contributions}
The proposed method can be classified as a CFAO method. However, instead of assigning design variables to each element, we use a neural-network (NN) to capture the matrix topology, the  fiber-orientation and fiber-density.  The advantages of the proposed framework are: \begin{enumerate*}[label=\itshape\alph*\upshape)]
	\item NN representation is smooth and differentiable over the domain, \item Expressing all computations in an end-end allows for automated sensitivity analysis \item one can extract continuous long fibers after optimization, \item the number of design variables is relatively small, and \item using a gradient based optimization, the three design variables can be simultaneously optimized. 
\end{enumerate*}

\begin{table}[h!]
	\caption{Comparison with recent literature}
	\begin{center}
		\begin{tabular}{  l | C{14mm} C{14mm} C{14mm}  }
			\hline \hline
			
			Reference & Matrix Topology & Fiber Density & Fiber Orientation   \\ \hline
			\cite{saketh20213PhaseFiberTO}, \cite{Chandrasekhar2020Fiber}, \cite{Stegmann2005}, \cite{brampton201} & $\checkmark$ & $\times $ & $\checkmark $  \\ \hline
			
			\cite{lee2018topology}, \cite{papapetrou2020stiffness}& $\checkmark$ & $\checkmark $ & $\times $  \\ \hline
			
			\cite{li2021full}, \cite{Steuben2016CADImplicitSlicing}, \cite{pedersen1991thickness} & $\times$ & $\times $ & $\checkmark $  \\ \hline

			Proposed  & $\checkmark $ & $\checkmark $ & $\checkmark$ \\ \hline
		\end{tabular}
	\end{center}
	\label{table:recentLitComparison}
\end{table}

\section{Proposed Method}
\label{sec:method}

\begin{figure*}[]
	\centering
	\begin{subfigure}[b]{0.5\linewidth}
		\centering
		\includegraphics[width=0.8\linewidth]{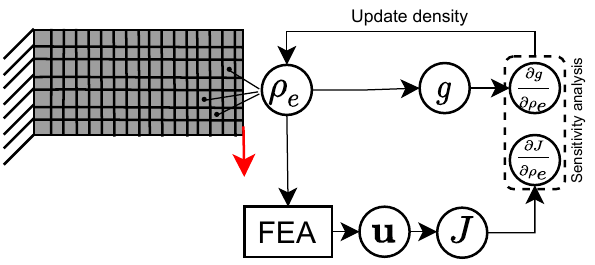}
		\caption{Single material mesh-based TO using SIMP}
		\label{fig:SIMP}
	\end{subfigure}
	\begin{subfigure}[b]{0.49\linewidth}
		\centering
		\includegraphics[width=\linewidth]{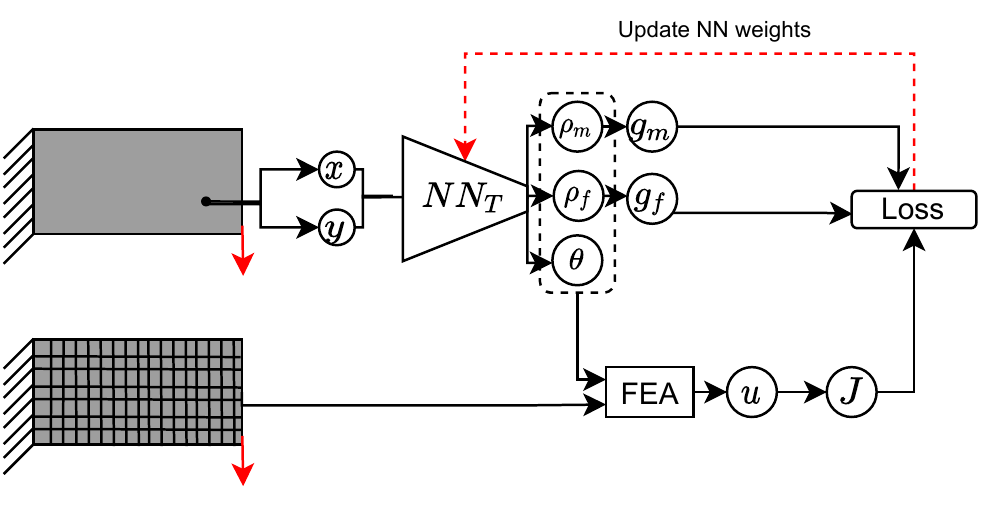}
		\caption{Optimization loop of the proposed FRC-TOuNN method}
		\label{fig:method_fiberOptimization}
	\end{subfigure}
	\caption{The underlying representation and design variables for typical single-material SIMP is the density at discrete finite elements.This coupling of representations for analysis and design can pose limitations in fidelity of design features. Also, the analytic expressions for sensitivity formulation must be found beforehand, which can be tedious and challenging. On the other hand, our proposed FRC-TOuNN approach decouples the representation for design (NN weights) from that of FEA (discrete elements), which enables geometric queries at high resolution (e.g., fiber orientation) while leveraging the efficiency of FEA. Further, our framework enables end-to-end automatic differentiability, where we only need to define a loss function without manually deriving sensitivity expressions.}
	\label{fig:method_comparison}
\end{figure*}

In this section, we discuss the proposed NN-based TO framework to optimize functionally graded FRC structures (FRC-TOuNN). \Cref{fig:method_comparison} illustrates the difference between the typical SIMP approach and the proposed method.  

In a typical single-material SIMP (\Cref{fig:SIMP}), the design is represented by the pseudo-density values at discrete finite elements $0 \le \rho_e \le 1$. The same discretization is used for FEA to evaluate the field variable ($\bm{u}$) and correspondingly the performance objective function (J) and other constraints (g). Given the analytical expressions for gradients with respect to the pseudo-density design variables, the elemental sensitivity values are computed. Then, by relying on the standard SIMP penalization technique \cite{bendsoe2013topology}, the density values are updated using gradient-based methods (e.g., optimality criteria \cite{berke1987OptimalityCriteria} or method of moving asymptotes \cite{svanberg1987MMA}).

This coupling of representations for analysis and design can pose limitations in the fidelity and accuracy of design features across different scales. Further, deriving the analytical expressions for sensitivity formulation can be tedious and challenging. The proposed FRC-TOuNN approach decouples the representation for design (NN weights) from that of FEA (discrete elements), which enables geometric queries at high resolution (e.g., fiber orientation) while leveraging the efficiency of FEA.  In other words, the design variables (here comprising of matrix density, fiber density, and fiber orientation) are captured externally using a NN's activation functions and weights and is independent of the underlying mesh. Thus, by relying on an appropriate material model (\Cref{sec:method_material}) and a loss function reflecting the objectives and constraints problem, the design is optimized using the NN’s built-in optimizer as depicted in \Cref{fig:method_fiberOptimization}.

\subsection{Design Variables}
\label{sec:method_overview}

We will assume that a design domain with loads and restraints (boundary conditions) has been prescribed (for example, see \Cref{fig:domain_topOpt_problem}). The objective here is to simultaneously optimize the (matrix) topology, fiber density, and fiber orientation to minimize an objective  while satisfying volume constraints on the matrix and fiber. To achieve this, we introduce three spatially varying  variables:  matrix density, $0 \le \rho_m \le 1$, fiber density, $0 \le \rho_f  \le 1$, and orientation, $-\pi /2 \le \theta \le \pi /2 $. These are collectively represented as  $\zeta(\bm{x}) = \{\rho_m(\bm{x}), \rho_f(\bm{x}), \theta(\bm{x})\}$. The matrix density, $\rho_m$, controls the overall topology, i.e.,  $\rho_m =0$ implies the matrix  is absent, while  $\rho_m =1$ implies the matrix is present. Intermediate densities will be discouraged through penalization as discussed later on. The fiber density, $\rho_f$, controls the concentration fibers; intermediate values of the fiber density are permitted. While we theoretically allow for the presence of fibers even if the base matrix is absent, this will not occur in practice due to an imposed volume control on the fiber volume. Finally, the fiber orientation, $ \theta$, allows for continuously varying orientations of fibers. 

\subsection{Optimization Problem}
\label{sec:optProblem}

To pose a valid  FRC optimization problem, we impose two-volume constraints. First, we impose a desired matrix volume fraction via $0 < V_m^* \le 1 $. This controls the overall topology of the design. Next, we control the fraction of this matrix that must be filled with fibers via $0 \le r_f^* \le 1 $. Consequently, one can pose the optimization problem as:

\begin{subequations}
	\label{eq:optimization_base_Eqn}
	\begin{align}
		& \underset{\bm{\rho_m}, \bm{\rho_f}, \bm{\theta}}{\text{minimize}}
		& &J(\bm{\rho_m}, \bm{\rho_f}, \bm{\theta}) \label{eq:optimization_base_objective}\\
		& \text{subject to}
		& & \bm{K}(\bm{\rho_m}, \bm{\rho_f}, \bm{\theta})\bm{u} = \bm{f}\label{eq:optimization_base_govnEq}\\
		& & & g_m (\bm{\rho_m})  \coloneqq \frac{\sum\limits_e \rho_m(\bm{x}_e) v_e}{V_m^*\sum\limits_e v_e} - 1 \leq 0  \label{eq:optimization_volConsMatrix}\\
		& & & g_f (\bm{\rho_f}) \coloneqq \frac{\sum\limits_e \rho_f(\bm{x}_e) v_e}{r_f^* V_m^*\sum\limits_e v_e} - 1  \leq 0 \label{eq:optimization_volConsFiber} \\
		& & & 0 \leq \rho_m, \rho_f \leq 1 \; , \;  -\frac{\pi}{2} \leq \theta \leq \frac{\pi}{2} \label{eq:optimization_boundConstraints}
	\end{align}
\end{subequations}

where $J$ is the design objective (compliance), and $v_e$ is the area of an element. Observe that with $r_f \leq 1$, the amount of fiber never exceeds the amount of base matrix. Further, under the optimal scenario, we expect the fiber density at a location to go to zero when the matrix density at that location goes to zero. No additional constraint is required/imposed.
\subsection{Neural Network  Representation}
\label{sec:method_NNRepresentation}

Typically, the  design variables in FRC optimization are captured via an underlying mesh. However, in this work, for reasons stated earlier,  the variables $\zeta(\bm{x}) = \{\rho_m(\bm{x}), \rho_f(\bm{x}), \theta(\bm{x})\}$ are represented  via a fully connected feed-forward NN \cite{Goodfellow2016_deepLearning} ($NN_T$), The NN architecture is as follows  (see \Cref{fig:NNlayers}):

\begin{figure}[]
	\begin{center}
		\includegraphics[scale=1.2,trim={0 0 0 0},clip]{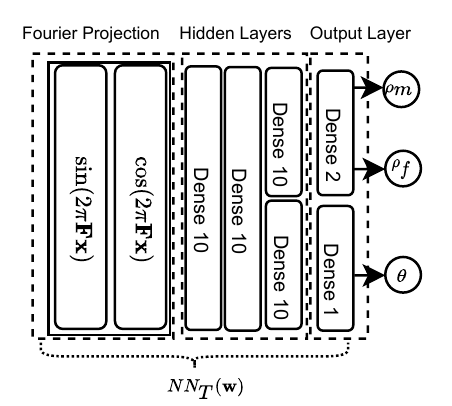}%
		\caption{Neural network layers for the $NN_T$ block of \Cref{fig:method_fiberOptimization}.}
		\label{fig:NNlayers}
	\end{center}
\end{figure}

\begin{enumerate}
	\item \textbf{Input Layer}: The input to the NN are points $\bm{x} \in \mathbf{R}^2$ within the domain.
	
	\item \textbf{Fourier Embedding}: The sampled points from the Euclidean input domain are first passed through a frequency space.  As detailed in \cite{tancik2020fourierNN}, \cite{Rahaman2019NNSpectralBias} implicit coordinate-based NN's perform poorly in capturing finer features. Following along \cite{chandrasekhar2021length}, we incorporate a Fourier projection layer that acts as an approximate length scale filter. In addition, the projection helps in faster and stable convergence of the optimizer. 
	\item \textbf{Hidden Layers}: The hidden layers  consist of a series of Swish ($x.sigmoid(x)$, \cite{ramachandran2017SwishActivation}) activated dense layers. To sufficiently demarcate the representation of the densities from the orientation, we introduce a fork in the hidden layers. In particular, we present two common layers, the outer layer interfacing with the Fourier projection layer. Further, the deeper common layer acts as an interface to two sub-layers with no interconnection among themselves. These two sub-layers individually connect to the densities and orientation outputs respectively.
	
	\item \textbf{Output Layer}: The output layer consists of three neurons corresponding to $\zeta = \{\rho_m, \rho_f, \theta\}$. The output layer is activated via a Sigmoid function $\sigma(\cdot)$ to ensure the outputs are in the range $(0,1)$. Further, the orientation is transformed as $\theta \leftarrow -\frac{\pi}{2} + \pi \sigma(\theta)$. 
\end{enumerate}

Thus the three design variables are now spatially varying and are controlled by the weights and bias of the NN. The weights and bias, denoted by the $\bm{w}$, now become the primary design variables, i.e., $ \rho_m(x,y; \bm{w})$ and so on. The strategy, therefore, is to optimize the matrix and fiber via the NN weights $\bm{w}$. Re-expressing the optimization problem in \Cref{eq:optimization_base_Eqn_NN} with the NN weights as design variables, we have,
\begin{subequations}
	\label{eq:optimization_base_Eqn_NN}
	\begin{align}
		& \underset{\bm{w}}{\text{minimize}}
		& &J(\bm{w}) \label{eq:optimization_base_objective_NN}\\
		& \text{subject to}
		& & \bm{K}(\bm{w})\bm{u} = \bm{f}\label{eq:optimization_base_govnEq_NN}\\
		& & & g_m (\bm{w})  \coloneqq \frac{\sum\limits_e \rho_m(\bm{x}_e) v_e}{V_m^*\sum\limits_e v_e} - 1 \leq 0  \label{eq:optimization_volConsMatrix_NN}\\
		& & & g_f (\bm{w}) \coloneqq \frac{\sum\limits_e \rho_f(\bm{x}_e) v_e}{r_f^* V_m^*\sum\limits_e v_e} - 1  \leq 0 \label{eq:optimization_volConsFiber_NN} \\
		& & & 0 \leq \rho_m, \rho_f \leq 1 \; , \;  -\frac{\pi}{2} \leq \theta \leq \frac{\pi}{2} \label{eq:optimization_boundConstraints_NN}
	\end{align}
\end{subequations}
\subsection{Material Model}
\label{sec:method_material}

We now discuss the underlying material model. The matrix is characterized by its elastic modulus $E_m$ and Poisson's ratio $\nu_m$; this yields the matrix elasticity tensor $ [D_m^0]$ \cite{chawla2012compositeTextbook}. The fiber is characterized by the elastic modulus along two axis $E^{\parallel}_f$, $E^{\perp}_f$, Poisson's ratio $\nu_f$ and shear modulus $G_f$ (assumed to be independent of Young's modulus); this yields the fiber elasticity tensor $ [D_f^0]$ \cite{chawla2012compositeTextbook}, when the fiber is oriented along the $x$-axis. Now given an arbitrary orientation $\theta (\bm{x})$, the fiber elasticity tensor is transformed as follows:
\begin{align}
	[D_f] = [\mathcal{T}_1]^{-1} [D^0_f] [\mathcal{T}_2]  
	\label{eq:D_matrix_net_fiber_matrix}
\end{align}
where $[\mathcal{T}_1]$ and $[\mathcal{T}_2]$ are the  transformation matrices:
\begin{equation}
	[\mathcal{T}_1] = \begin{bmatrix}
		m^2 & n^2 & mn \\
		n^2 & m^2 & -2mn \\
		-mn & mn & m^2 - n^2 \\
	\end{bmatrix} \; , \; 	[\mathcal{T}_2] = \begin{bmatrix}
		m^2 & n^2 & mn \\
		n^2 & m^2 & -mn \\
		-2mn & 2mn & m^2 - n^2 \\
	\end{bmatrix} 
	\label{eq:rotationTransformationMatrices}
\end{equation}
where $m = \cos \theta$ and $n = \sin \theta$ (the reader is referred to \cite{chawla2012compositeTextbook} for details). Note that the angle $\theta$, and therefore, the transformation matrices are spatially varying, but the dependency on $\bm{x}$ has been suppressed for readability. Finally,  given a matrix density, $\rho_m$, and a fiber density, $\rho_f$, the effective elasticity tensor $[D]$ at any point $\bm{x}$ can be expressed as:

\begin{align}
	[D(\bm{x})] = \rho_m^p\bigg( &\rho_f [\mathcal{T}_1 ]^{-1} [D^0_f] [\mathcal{T}_2]  +  ( 1-\rho_f)[D^0_m] \bigg)
	\label{eq:D_matrix_net_fiber_matrix_final}
\end{align}
As mentioned earlier, the matrix density, $\rho_m$, is penalized using  the SIMP penalty parameter $p$ to drive the design towards a $0/1$ topology. On the other hand, the fiber density, $\rho_f$, is allowed to take intermediate values.

\subsection{Finite Element Analysis}
\label{sec:method_FEA}

We will use conventional finite element analysis (FEA) as part of the optimization engine. Here, we use a structured quadrilateral mesh due to its simplicity (see \Cref{fig:method_fiberOptimization}).

One can now evaluate the element stiffness matrix as

\begin{equation}
	[K_e] = \int\limits_{\Omega_e} [\nabla N_e]^T[D(\bm{x}_e)][\nabla N_e] d \Omega_e
	\label{eq:K_matrix_FE expression}
\end{equation}

where $[\nabla N_e]$ is the gradient of the shape matrix, and $[D(\bm{x}_e)]$ is the elasticity tensor evaluated at the center of the element. Since the elasticity tensor varies spatially, the element stiffness matrix must be computed for each element, and for each step of the optimization process. This can be computationally very expensive.  We, therefore, exploit the concept of \emph {template stiffness matrices} to reduce the computational effort.

Let

\begin{equation}
	[D(\bm{x_e})] = \begin{bmatrix}
		D_{11} & D_{12}  & D_{13}  \\
		D_{21}  & D_{22}  & D_{23}  \\
		D_{31}  & D_{32}  & D_{33}  \\
	\end{bmatrix}
\end{equation}
One can therefore express the stiffness matrix as follows

\begin{equation}
	\begin{aligned}
		[K_e] &=  D_{11} [\hat{K}^{1}] + D_{22}  [\hat{K}^{2}] +  D_{33}  [\hat{K}^{3}] \\&+ D_{12}  [\hat{K}^{4}] +  D_{13}  [\hat{K}^{5}] + D_{23}  [\hat{K}^{6}]
		\label{eq:stiffnessElem_D_times_Ktemplate}
	\end{aligned}
\end{equation}
where,

\begin{equation}
	[\hat{K}^i] = \int\limits_{\Omega_e} [\nabla N_e] ^T[\hat{D}^i][\nabla N_e]  d \Omega_e, \quad i = 1,2, \ldots 6
	\label{eq:K_matrix_hat_1}
\end{equation}
and
\begin{align}
	[\hat{D}^1] = \begin{bmatrix}
		1 & 0 & 0 \\
		0 & 0 & 0 \\
		0 & 0 & 0 \\
	\end{bmatrix} \;  [\hat{D}^2] = \begin{bmatrix}
		0 & 0 & 0 \\
		0 & 1 & 0 \\
		0 & 0 & 0 \\
	\end{bmatrix} \quad \; [\hat{D}^3] = \begin{bmatrix}
		0 & 0 & 0 \\
		0 & 0 & 0 \\
		0 & 0 & 1 \\
	\end{bmatrix}  \nonumber\\ 
	[\hat{D}^4] = \begin{bmatrix}
		0 & 1 & 0 \\
		1 & 0 & 0 \\
		0 & 0 & 0 \\
	\end{bmatrix}  \; 	[\hat{D}^5] = \begin{bmatrix}
		0 & 0 & 1 \\
		0 & 0 & 0 \\
		1 & 0 & 0 \\
	\end{bmatrix} \quad  \; 	[\hat{D}^6] = \begin{bmatrix}
		0 & 0 & 0 \\
		0 & 0 & 1 \\
		0 & 1 & 0 \\
	\end{bmatrix}
\end{align}

Observe that matrices $ [\hat{K}^i]$ need to be computed only once at the start of the optimization. Then, during optimization, the element stiffness matrices are computed efficiently via Equation \ref{eq:stiffnessElem_D_times_Ktemplate}, followed by the assembly of the global stiffness matrix $\bm{K}$. During each step of the optimization problem, we  solve the structural problem
\begin{align}
	\bm{K}\bm{u} = \bm{f}
	\label{eq:FEA}
\end{align}
where $\bm{f}$ is the imposed load,  and $\bm{u}$ is the displacement field.

\subsection{Loss Function}
\label{sec:method_LossFunction}
NNs are designed to minimize an unconstrained function. We therefore convert the constrained optimization problem in  \Cref{eq:optimization_base_Eqn_NN} to an  unconstrained loss function via a simple penalty formulation \cite{nocedal2006numericalOptimization}:
\begin{equation}
	L(\bm{w}) = \frac{J(\bm{w}) }{J_0} + \alpha_m (g_m(\bm{w})) ^2 + \alpha_f (g_f(\bm{w}))^2
	\label{eq:lossFunction}
\end{equation}
where $J_0$ is the initial compliance, used here for scaling, and $\alpha_m$, $\alpha_f$ are the penalty parameters. While they can vary independently, we treat them to be the same, i.e., $\alpha_m = \alpha_f = \alpha$ in our experiments; please see Section \ref {sec:numExpts} for additional details. The equality constraint \Cref{eq:optimization_base_govnEq_NN} is satisfied when solving for the displacement during FEA (\Cref{sec:method_FEA}) while box constraints of \Cref{eq:optimization_boundConstraints_NN} are satisfied automatically by the output of the NN (\Cref{sec:method_NNRepresentation}). The optimization scheme is illustrated in \Cref{fig:method_fiberOptimization}.

\begin{figure*}[]
	\begin{center}
		\includegraphics[scale=1,trim={0 0 0 0},clip]{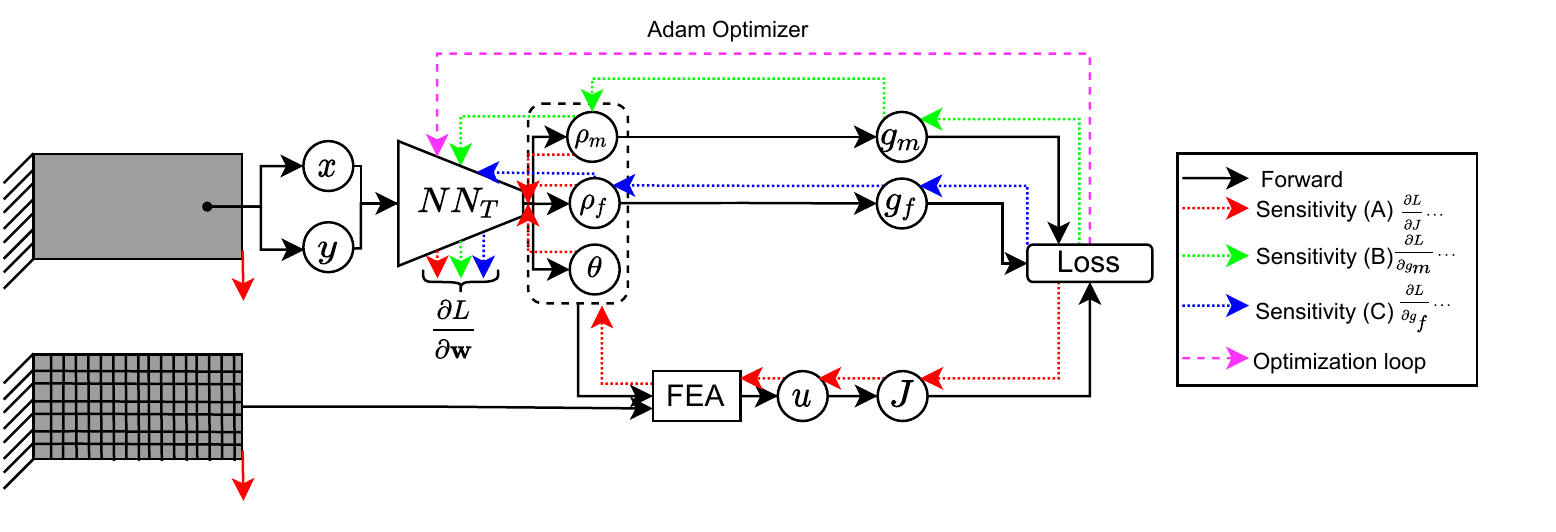}%
		\caption{Forward and Backward computations}
		\label{fig:fwdAndBwdComputation}
	\end{center}
\end{figure*}
\subsection{Sensitivity Analysis}
\label{sec:method_sensAnalysis}
A critical ingredient for the update schemes during optimization is the sensitivity, i.e., derivative, of the loss function (\Cref{eq:lossFunction}) with respect to the weights of the NN. Typically the sensitivity analysis is carried out manually. This can be laborious and error-prone, especially for non-trivial objectives. Here, by expressing all our computations including the FEA in PyTorch \cite{NEURIPS2019_9015_pyTorch}, we rely on the automatic differentiation (AD) capabilities of NNs to completely automate this step \cite{Chandrasekhar2021AuTO}. AD performs a systematic derivation of the sensitivity using the chain-rule. We express the broader derivative terms in \Cref{eq:partial_sens_loss}. The forward and sensitivity computations are schematically depicted in \Cref{fig:fwdAndBwdComputation}.

\begin{equation}
	\begin{aligned}
		\frac{\partial L}{\partial \bm{w}} &=\overbrace{\frac{\partial L}{\partial J} \frac{\partial J}{\partial \bm{u}} \bigg[ \frac{\partial \bm{u}}{\partial \bm{\theta}} \frac{\partial \bm{\theta}}{\partial \bm{w}} + \frac{\partial \bm{u}}{\partial \bm{\rho_f}} \frac{\partial \bm{\rho_f}}{\partial \bm{w}} +  \frac{\partial \bm{u}}{\partial \bm{\rho_m}} \frac{\partial \bm{\rho_m}}{\partial \bm{w}} \bigg]}^{A} \\  
		&+ \overbrace{\frac{\partial L}{\partial g_m} \frac{\partial g_m}{\partial \bm{\rho_m}}\frac{\partial \bm{\rho_m}}{\partial \bm{w}}}^{B} + \overbrace{\frac{\partial L}{\partial g_f} \frac{\partial g_f}{\partial \bm{\rho_f}} \frac{\partial \bm{\rho_f}}{\partial \bm{w}}}^{C}
		\label{eq:partial_sens_loss}
	\end{aligned}
\end{equation}


\subsection{Post Processing}
\label{sec:expts_postProcessing}
\begin{figure}[]
	\begin{center}
		\includegraphics[scale=0.25,trim={0 0 0 0},clip]{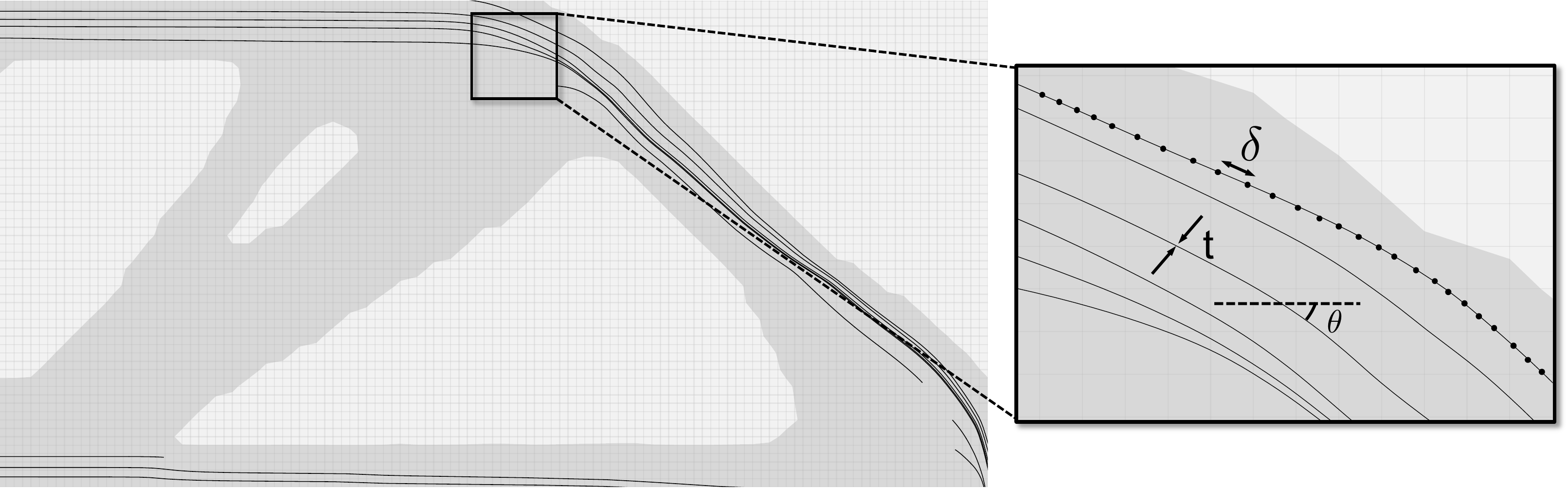}%
		\caption{Post processing to obtain continuous fibers}
		\label{fig:postProcessingGraphic}
	\end{center}
\end{figure}

In this section, we address how one may interpret the output from the neural network post optimization to obtain continuous fiber tracks, where each track is represented by a set of ordered points. The global NN-representation  allows us to query the relevant quantities  at any location in the domain, i.e., $\zeta(\vec{x}) = NN(\vec{x}, \bm{w}^*)$. We assume that the fiber thickness ($t$) and step size ($\delta$) are specified.  In the  post processing algorithm below, we generate fibers and continuously update the fiber density $\rho_{f}^{actual}$ to match the desired fiber-density $\rho_{f}$ as described below. 
	\begin{enumerate}
		\item $\rho_{f}^{actual}$  is initialized to zero  for all elements.
		\item We start at any non-void element where  $\rho_{f}^{actual}$  is less than the desired  $\rho_{f}$. If no such element exists, the post-processing algorithm stops.
		\item A fiber path is initiated at this element and is  appended to a list;  the quantity  $\rho_{f}^{actual}$ is updated for that element j (as the volume fraction of fiber within the element). 
		\item Using the orientation $\theta$ and step-size  $\delta$ at that element we take a step (in both directions, in two passes) to the next element. 
		\item If: (a) the next element exists, and (b) $\rho_{f}^{actual}$ is less than $\rho_{f}$, we continue to along the path, else we terminate this fiber-path and continue to step-2.  
	\end{enumerate}
	For instance \Cref{fig:postProcessingGraphic} illustrates a few fibers extracted using the proposed post processing algorithm. Note that if the fiber thickness is much smaller than the element size, and if the desired fiber density is high, then multiple fibers can pass through an element. 

\subsection{Optimization Algorithm}

\label{sec:method_OptimizationAlgorithm}

The complete algorithm of the proposed framework is summarized in \Cref{alg:FRC-TOuNN}, and schematically depicted in \Cref{fig:method_fiberOptimization}.

\begin{algorithm}[!ht]
	\caption{FRC-TOuNN}
	\label{alg:FRC-TOuNN}
	\begin{algorithmic}[1]
		\Procedure{TopOpt}{$\Omega^0$ , BC,   $V_m^*, r_f^*$}
		
		\Comment{Inputs}
		
		\State $\Omega^0 \rightarrow \Omega^0_h$ \Comment{Discretize domain for FE} \label{alg:domainDiscretize}
		
		\State $\Omega^0_h \rightarrow [\hat{K}^{i}]$ \Comment{Compute stiff.  templates \cref{eq:K_matrix_hat_1}} \label{alg:stiffnessTemplates}
		
		\State $\bm{x} = \{x_e,y_e\}_{e \in \Omega^0_h} \quad \bm{x} \in \mathbb{R}^{n_e \times 2}$ \Comment{elem centers; NN input} \label{alg:elemCenterComp}
		
		\State $J_0 \leftarrow FEA(\bm{\rho}_m = V_m^*, \bm{\rho}_f = r_f^*, \Omega^0_h$, BC) \Comment {init obj} \label{alg:initComplianceComp}
		
		\State $F \sim \mathcal{U}(\frac{h}{l_{max}}, \frac{h}{l_{min}})$ \; ; $F \in \mathbb{R}^{2 \times n_f}$ \Comment{Uniformly sampled freq} \label{alg:fourierProjInit}
		
		\State $\bm{w} \sim \text{Xavier normal}$ \Comment{Init. NN weights \cite{glorot2010understanding}} \label{alg:initNNwts}
		
		\State  epoch = 0; $\alpha = \alpha_0$; $p = p_0$ \Comment{Initialization} \label{alg:initalizationParams}
		
		\Repeat \Comment{Optimization (Training)}
		
		\State $NN_T(\bm{x} ; \bm{w}) \rightarrow \bm{\zeta}(\bm{x})$ \Comment{Fwd prop through NN} \label{alg:fwdPropNN}
		
		\State $\bm{\zeta}(\bm{x}) \rightarrow [D(\bm{x})]$ \Comment{Elasticity tensor  \cref{eq:D_matrix_net_fiber_matrix_final}} \label{alg:elasticityTensorCompute}
		
		\State$([D(\bm{x})],  [\hat{K}^{i}]) \rightarrow [K(\bm{x})]$ \Comment{Elem stiff matrix \cref{eq:stiffnessElem_D_times_Ktemplate} } \label{alg:elemStiffnessCompute}
		
		\State $\bm{u}$ via solving $\bm{K}(\bm{\zeta})\bm{u} = \bm{f}$ \Comment{ FEA \cref{eq:optimization_base_govnEq}} \label{alg:feSolve}
		
		\State$([K], \bm{u}) \rightarrow J$ \Comment{Objective, \Cref{eq:optimization_base_objective}} \label{alg:objectiveComp}
		
		\State $\{\rho_m(\bm{x}), V_m^* \} \rightarrow g_m$ \Comment{Matrix vol cons, \cref{eq:optimization_volConsMatrix}} \label{alg:matrixVolCons}
		
		\State $ \{\rho_f(\bm{x}), r_f^* \} \rightarrow g_f$ \Comment{Fiber vol cons, \cref{eq:optimization_volConsFiber}} \label{alg:fiberVolCons}
		
		\State $(J, g_m, g_f) \rightarrow L$ \Comment{Loss from \Cref{eq:lossFunction}} \label{alg:lossCompute}
		
		\State $AD(L \leftarrow \bm{w}) \rightarrow \nabla L $ \Comment{sensitivity analysis} \label{alg:autoDiff}
		
		\State $w  +  \Delta w (\nabla L ) \rightarrow w $ \Comment{Adam optimizer \cite{Kingma2015ADAM} step}\label{alg:adamStep}
		
		\State $ \text{min} (\alpha _{max}, \alpha + \Delta \alpha)  \rightarrow \alpha$ \Comment {Increment $\alpha$} \label{alg:OptPenaltyUpdate}
		
		\State $  \text{min} (p_{max}, p + \Delta p ) \rightarrow p$ \Comment {Continuation } \label{alg:materialPenaltyUpdate}
		
		\State $\text{epoch}++$
		
		\Until{ $|| \Delta w || < \Delta w_c^*$ } \Comment{Check for convergence}
		
		\EndProcedure
	\end{algorithmic}
\end{algorithm}

\section{Numerical Experiments}
\label{sec:numExpts}

In this section, we conduct several experiments to illustrate the proposed framework. The implementation is in Python and uses the PyTorch library \cite{NEURIPS2019_9015_pyTorch}. The default parameters in the experiments are summarized in \Cref{table:defaultParameters}.  Through the experiments, we investigate the following.
\begin{enumerate}
	
	\item \textbf{Validation}: We benchmark results obtained via our method against published results \cite{saketh20213PhaseFiberTO}.
	
	\item \textbf{Convergence Study}:  Typical convergence plots of the objective and volume constraints are reported and explained.
	
	\item \textbf{Pareto Designs}: Trade-off in performance when optimizing for varying volume fractions of the fiber and matrix is reported. 
	
	\item \textbf{Computational Cost}: The cost to solve the optimization problem as a function of the number of elements used in FEA is studied. The cost division between the various components of the framework is summarized.
	
	\item \textbf{Compliant Mechanism}: By expressing our computations in an end-to-end differentiable form, we gain the ability to experiment with various objectives easily. We showcase this by designing  compliant mechanisms. 
	
	\item \textbf{Multi-material}: Fibrous composite structures often comprise regions made of different materials. We combine the present formulation with \cite{chandrasekhar2021multi} to obtain structures with  multi-material fibrous composites. 
	
\end{enumerate}
\begin{table}[!t]
	\caption{Simulation parameters}
	\begin{center}
		\begin{tabular}{  r | p{100mm}  }
			
			Parameter & Description and default value \\ \hline
			$E_m$, $\nu_m$ & Isotropic matrix Young's Modulus $E_m = 1 GPa$, Poisson's ratio $\nu = 0.3$ \\
			$E^{\parallel}_f$, $E^{\perp}_f$,$\nu_f$, $G_{f}$ & Orthotropic fiber inclusion material constants.\newline $E^{\parallel}_f = 4 GPa$, $E^{\perp}_f = 2 GPa$,$\nu_f = 0.3$, $G_{f} = 0.7 GPa$ \\
			$V_m^*$, $r_f^*$ & Allowed matrix volume fraction and fiber fraction\\
			$n_f$ & Number of frequencies in projection space = 150 \\
			$l_{min}, l_{max}$ & $l_{min} = 4$ and $l_{max} = 80$ to allow wide range of length scales. \\
			$\alpha_{max}$, $\Delta \alpha$ & Penalty method constraint with ($\alpha_{max}= 100$, $\Delta \alpha =\alpha_{0}= 0.05$) \\
			$p$ & SIMP penalty updated every iteration ($p_0 = 1 \; ; \; \Delta p = 0.02 \; ; \; p_{max} = 8$) \\
			lr & Adam Optimizer learning rate 0.01 \\
			nelx, nely & Number of FEA mesh elements of (60, 30)  along $x$ and $y$ respectively \\
			h & Size of element along $x = y = 1 \; mm$ \\
			$\Delta w_c^*$ & Convergence criteria requiring a change in norm of the weights to be less than $0.005$ \\
			
		\end{tabular}
	\end{center}
	
	\label{table:defaultParameters}
\end{table}
\subsection{Validation}
\label{sec:expts_validation}

\begin{figure}[!ht]
	\centering
	
	\begin{subfigure}[b]{0.8\linewidth}
		\centering
		\includegraphics[width=0.5\linewidth]{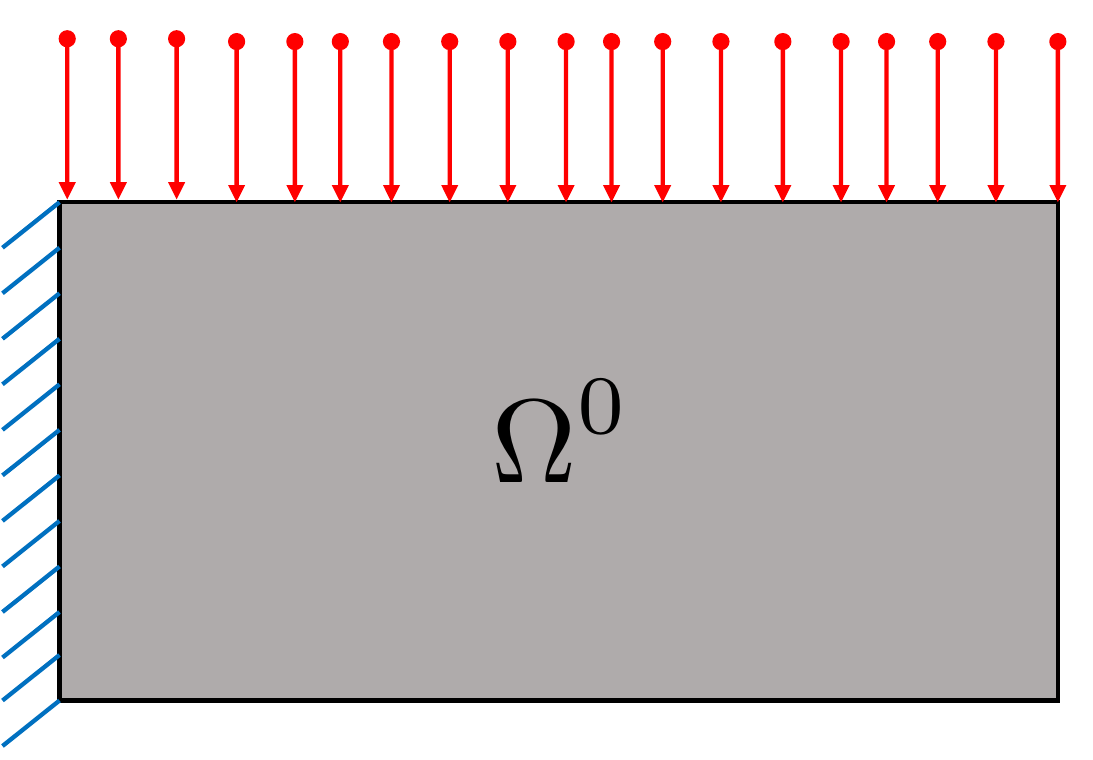}
		\caption{Top loaded beam}
	\end{subfigure}
	
	\begin{subfigure}[b]{0.8\linewidth}
		\centering
		\includegraphics[width=0.5\linewidth]{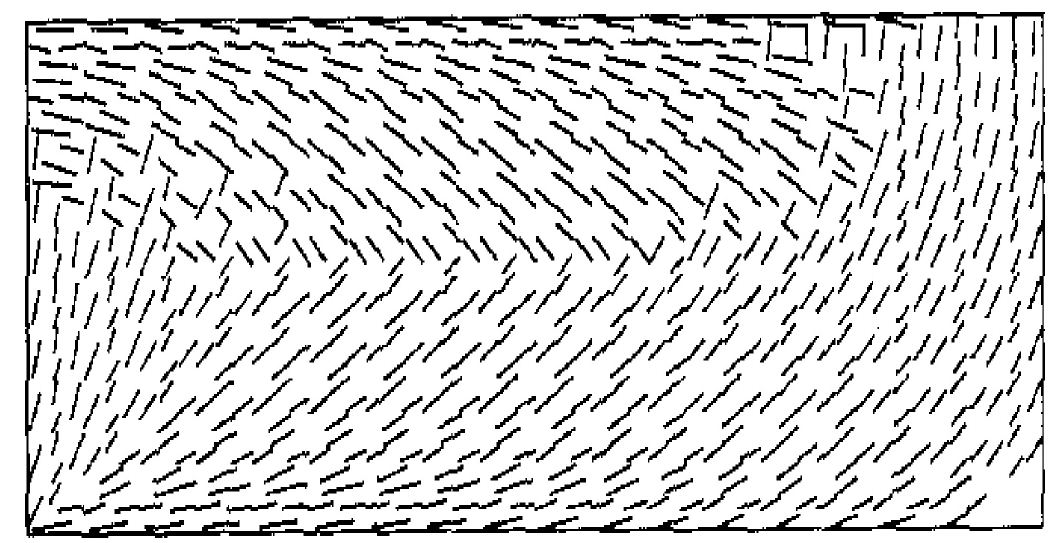}
		\caption{Result from \cite{pedersen1991thickness}}
	\end{subfigure}
	
	\begin{subfigure}[b]{0.8\linewidth}
		\centering
		\includegraphics[width=0.5\linewidth]{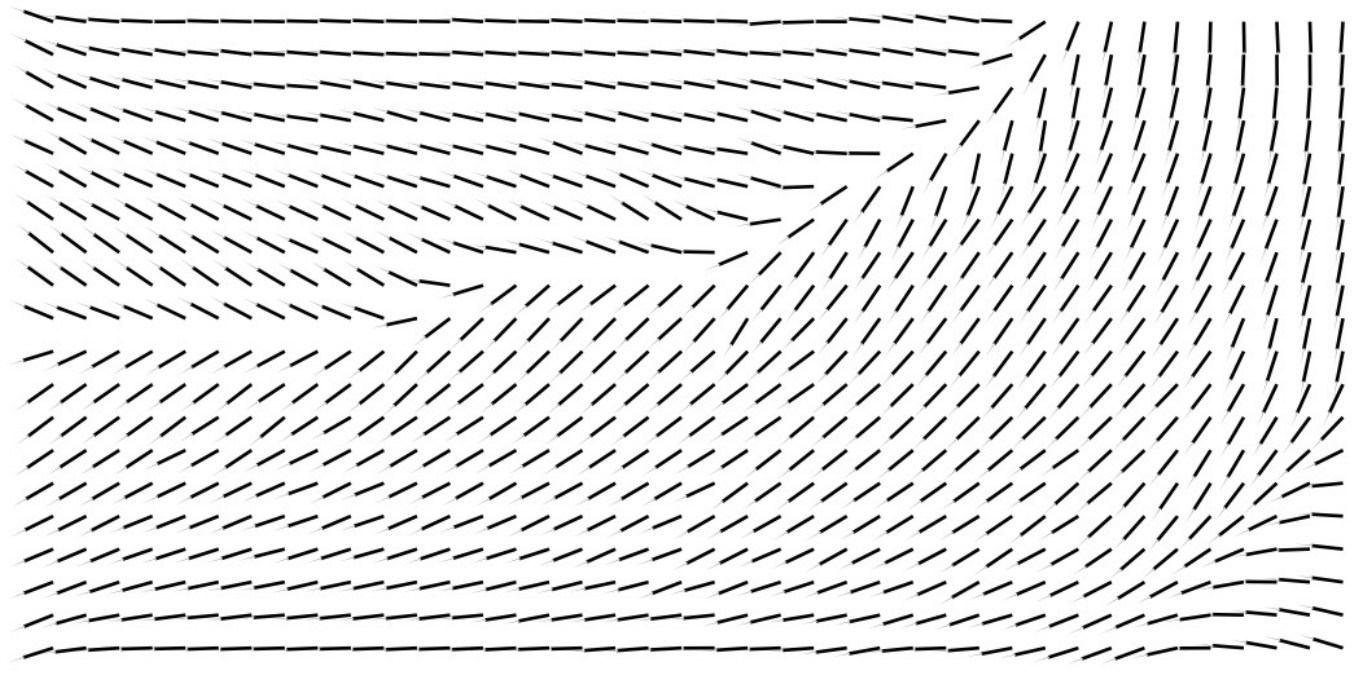}
		\caption{Proposed method}
	\end{subfigure}
	
	\begin{subfigure}[b]{0.8\linewidth}
		\centering
		\includegraphics[width=0.5\linewidth]{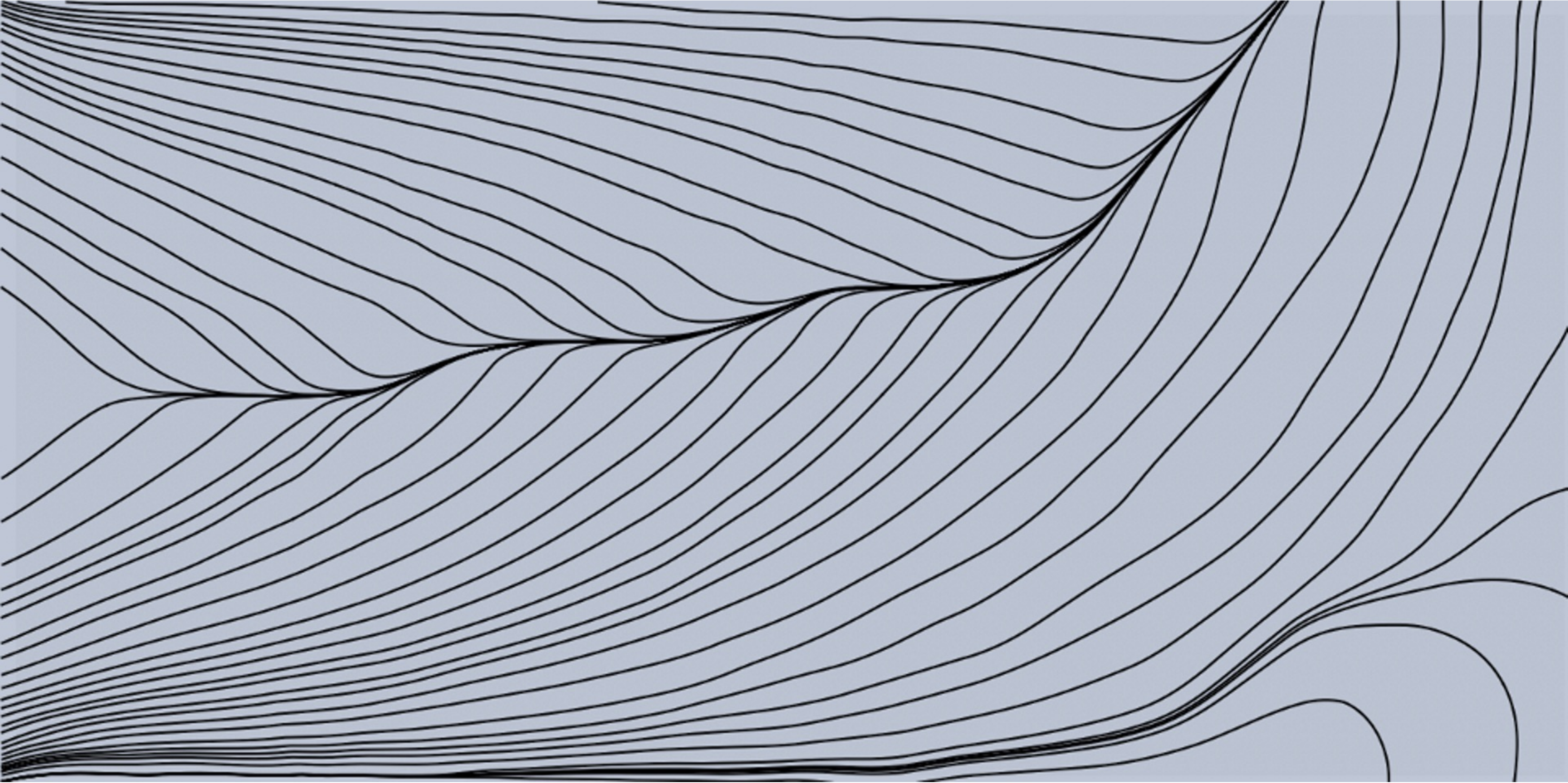}
		\caption{Extracted continuous fibers}
	\end{subfigure}

	\begin{subfigure}[b]{0.9\linewidth}
		\centering
		\includegraphics[width=0.5\linewidth]{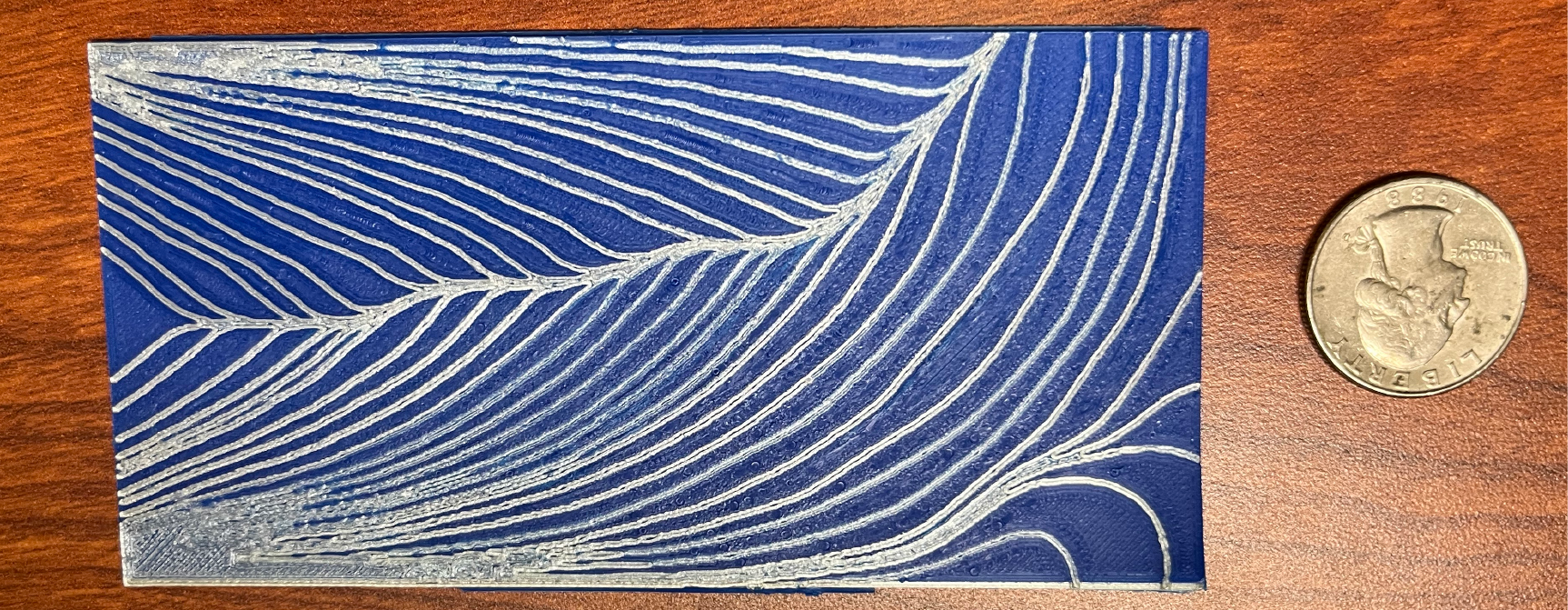}
		\caption{Printed FRC with continuous fibers}
	\end{subfigure}
	
	\caption{Comparison of result obtained for a top loaded beam between proposed method and \cite{pedersen1991thickness} }
	\label{fig:result_validation_topBeam_result}
\end{figure}

For validation, first, we replicate the FRC problem posed in  \cite{pedersen1991thickness} where only the fiber orientation of a beam loaded uniformly on its top edge (\Cref{fig:result_validation_topBeam_result}) is to be optimized. We set $V_m^* = r_f^* = 1$. We compare the results obtained  (\Cref{fig:result_validation_topBeam_result}(c)) with the reported topologies in \cite{pedersen1991thickness} (\Cref{fig:result_validation_topBeam_result}(b)). We observe that the reported orientations are comparable. \Cref{fig:result_validation_topBeam_result}(d) shows the post-processing result with continuous fibers and \Cref{fig:result_validation_topBeam_result}(e) the 3D printed part. We note that the prints presented in this work are purely representative and additional manufacturing constraints may need to be imposed to render the parts manufacturable by current fiber printers. 

\begin{figure*}[]
	\begin{center}
		\includegraphics[ scale=0.8,trim={0 0 0 0},clip]{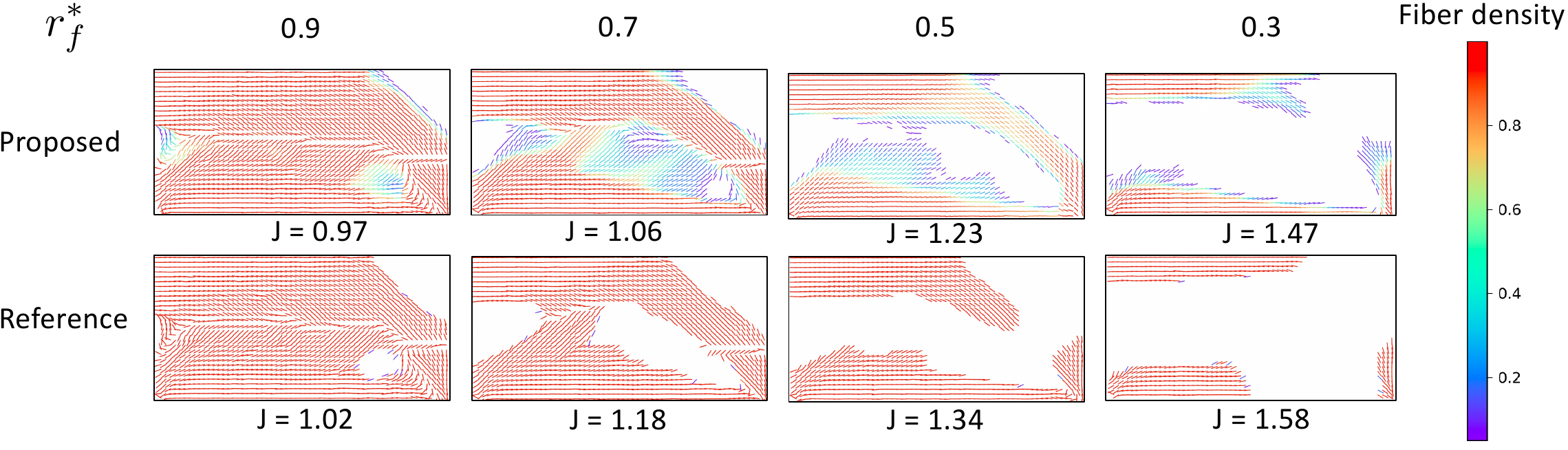}%
		\caption{Comparison of proposed method with \cite{saketh20213PhaseFiberTO} for various $r_f^*$ for the tip cantilever beam ($30 \times 60 \; cm^2$) example of \Cref{fig:domain_topOpt_problem}.}
		\label{fig:result_validation_2phase}
	\end{center}
\end{figure*}

\begin{figure}[!ht]
	\begin{subfigure}[b]{\linewidth}
		\centering
		\includegraphics[width=0.6\linewidth]{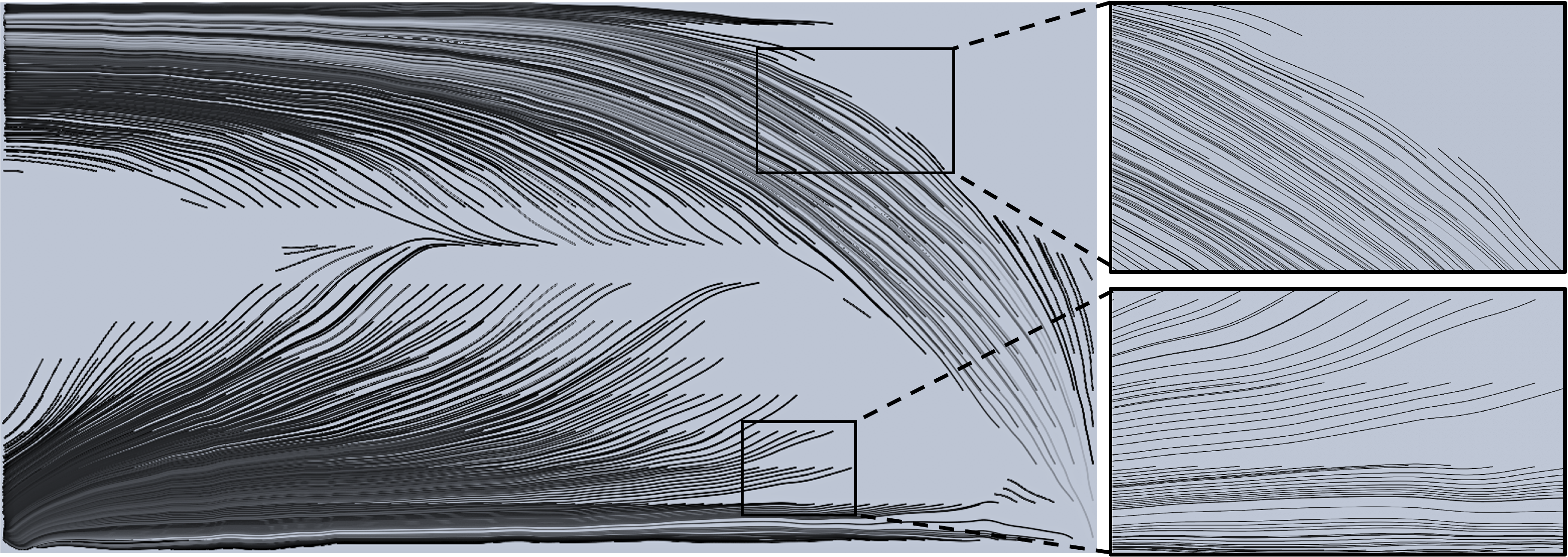}
		\caption{Generated long fibers with fiber thickness of 0.1 mm }
		\label{fig:fibers_tipCant_all_image}
	\end{subfigure}
	
	\begin{subfigure}[b]{\linewidth}
		\centering
		\includegraphics[width=0.5\linewidth]{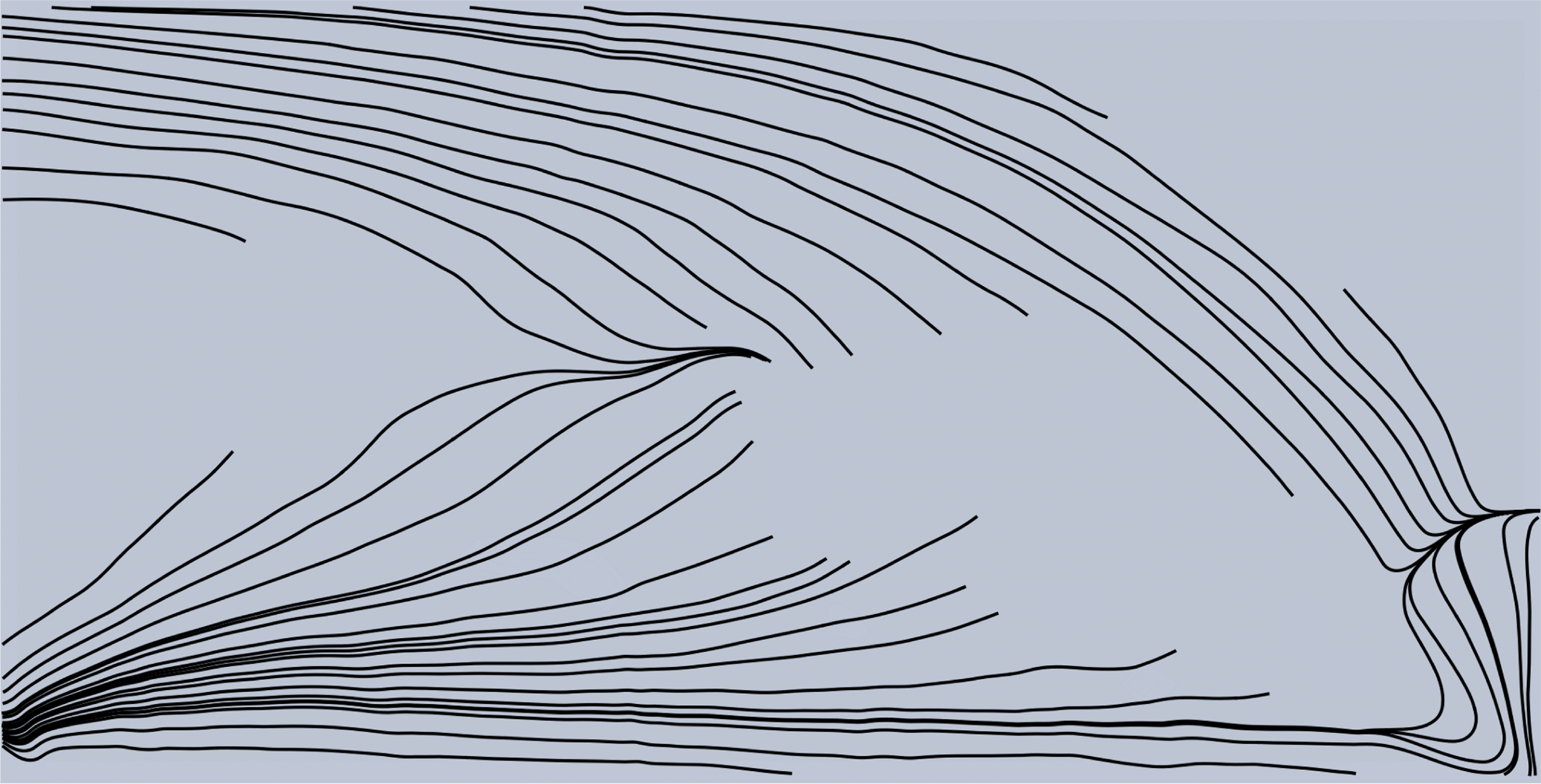}
		\caption{Generated long fibers with a thickness of 1mm. }
		\label{fig:tipCantPrintedImage}
	\end{subfigure}
	\hfill 
	\begin{subfigure}[b]{\linewidth}
		\centering
		\includegraphics[width=0.5\linewidth]{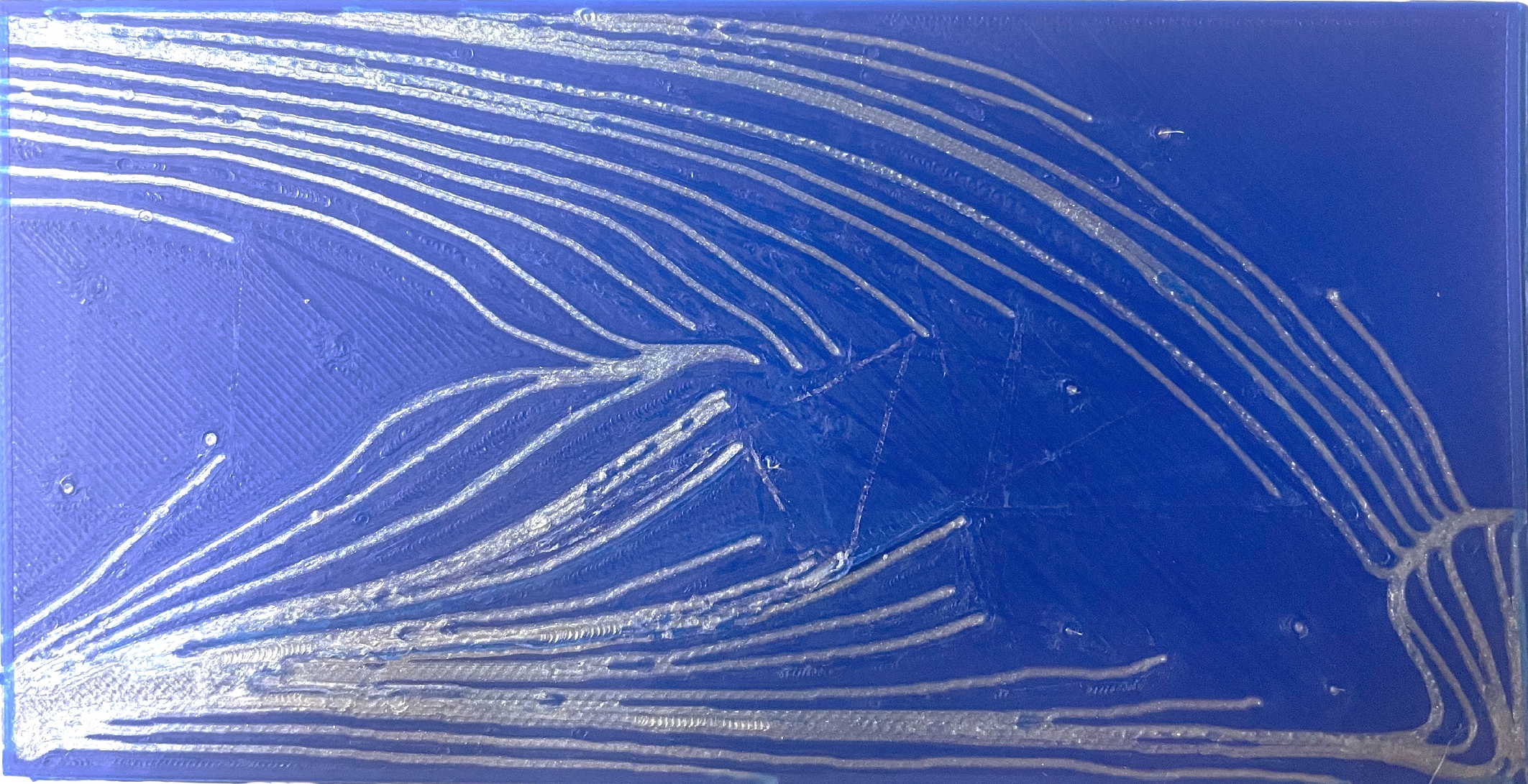}
		\caption{Printed FRC with continuous fibers}
		\label{fig:tipCantPrinted}
	\end{subfigure}
	\caption{Extracted continuous fiber design and print for a tip cantilever beam with $r_f^* = 0.5$}
\end{figure}

As a second validation experiment, we consider the results presented in \cite{saketh20213PhaseFiberTO}.  where the matrix was not optimized, and the fiber is  present or absent within the matrix, i.e., the concentration is either 1 or 0. The allowable fiber fraction $r_f^*$ was varied keeping $V_m^*=1$ and we compare the results with the proposed method and report the compliance and obtained topologies in \Cref{fig:result_validation_2phase}. \textit{We observe that the material gradation as obtained by the variable spacing/concentration of the fibers results in better designs with lower compliance values.} The result obtained for $r_f^*=0.5$ was post processed to obtain long continuous fibers (with a representative fiber thickness of 0.1mm) as illustrated in \Cref{fig:fibers_tipCant_all_image}. For illustration thick fibers (1mm) were generated and printed using an $\textit{UltiMaker 3}$ printer.

\subsection{Convergence}
\label{sec:convergence}
\begin{figure}[!ht]
	\begin{center}
		\includegraphics[scale=0.75,trim={0 0 0 0},clip]{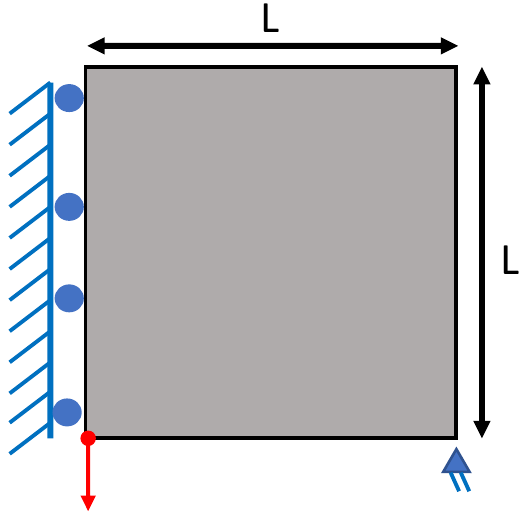}
		\caption{Symmetric half of Michell structure (L = 30 cm)}
		\label{fig:michellStructure}
	\end{center}
\end{figure}

\begin{figure}[!ht]
	\begin{center}
		\includegraphics[scale = 1.,trim={0 0 0 0},clip]{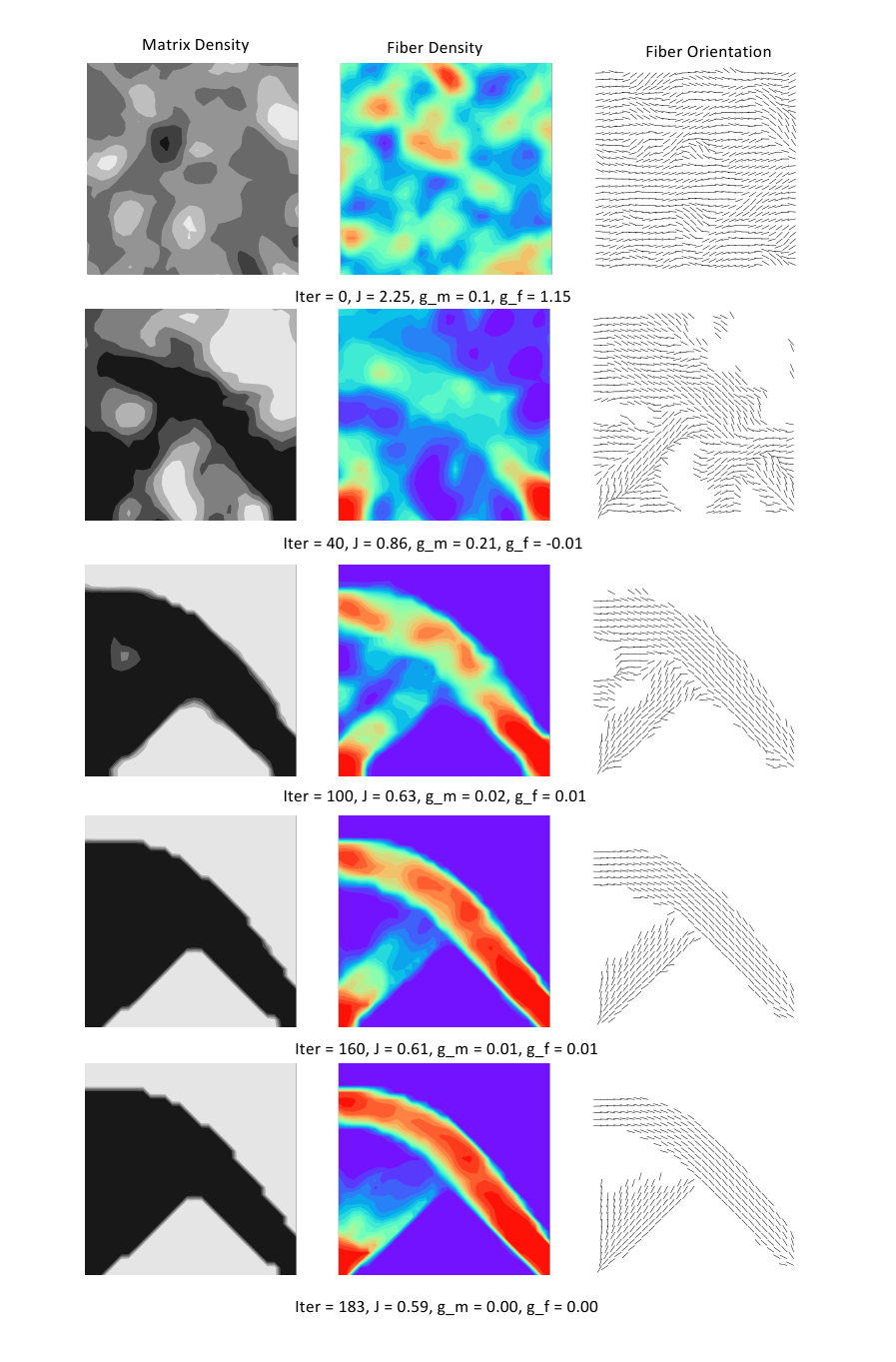}
		\caption{Typical snapshots during optimization.}
		\label{fig:result_convergence}
	\end{center}
\end{figure}

In this example, we illustrate the typical convergence of the proposed algorithm using the symmetric half of the Michell structure \Cref{fig:michellStructure} with $V_m^* = 0.5$ and $r_f^* = 0.5$. The compliance and the volume constraints of the matrix and fiber along with the matrix and fiber layout are illustrated at various instances in \Cref{fig:result_convergence}.  We observed a stable convergence using the simple penalty formulation.

\subsection{Pareto Designs}
\label{sec:result_Pareto}
\begin{figure}
	\begin{center}
		\includegraphics[scale = 0.4,trim={0 0 0 0},clip]{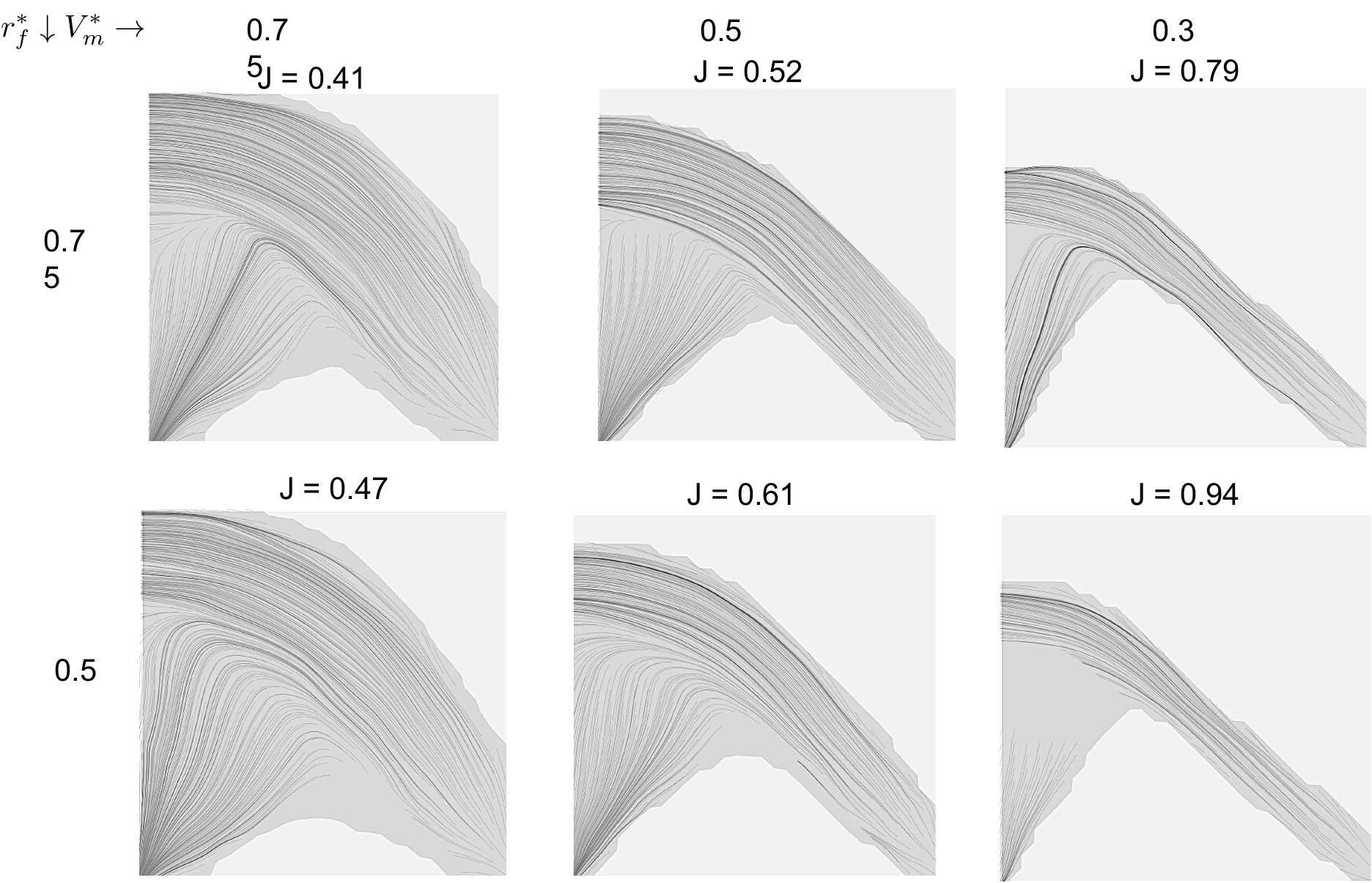}
		\caption{Varying allowed fiber and matrix fractions results in performance trade-offs.}
		\label{fig:result_paretoDesignFiberConc}
	\end{center}
\end{figure}

An important consideration during the design stage is exploring the Pareto-front to make judicious engineering decisions. Towards this goal, we once again consider the problem in \Cref{fig:michellStructure} and explore the compliance and topology for varying values of $r_f^*$ and $V_m^*$ in \Cref{fig:result_paretoDesignFiberConc}. As expected, we observe increasing compliance as we lower the allowed volume fractions.

\subsection{Manufacturing Considerations}
\label{sec:expts_manufCons}
\begin{figure}[]
	\begin{subfigure}[b]{\linewidth}
		\centering
		\includegraphics[width=0.5\linewidth]{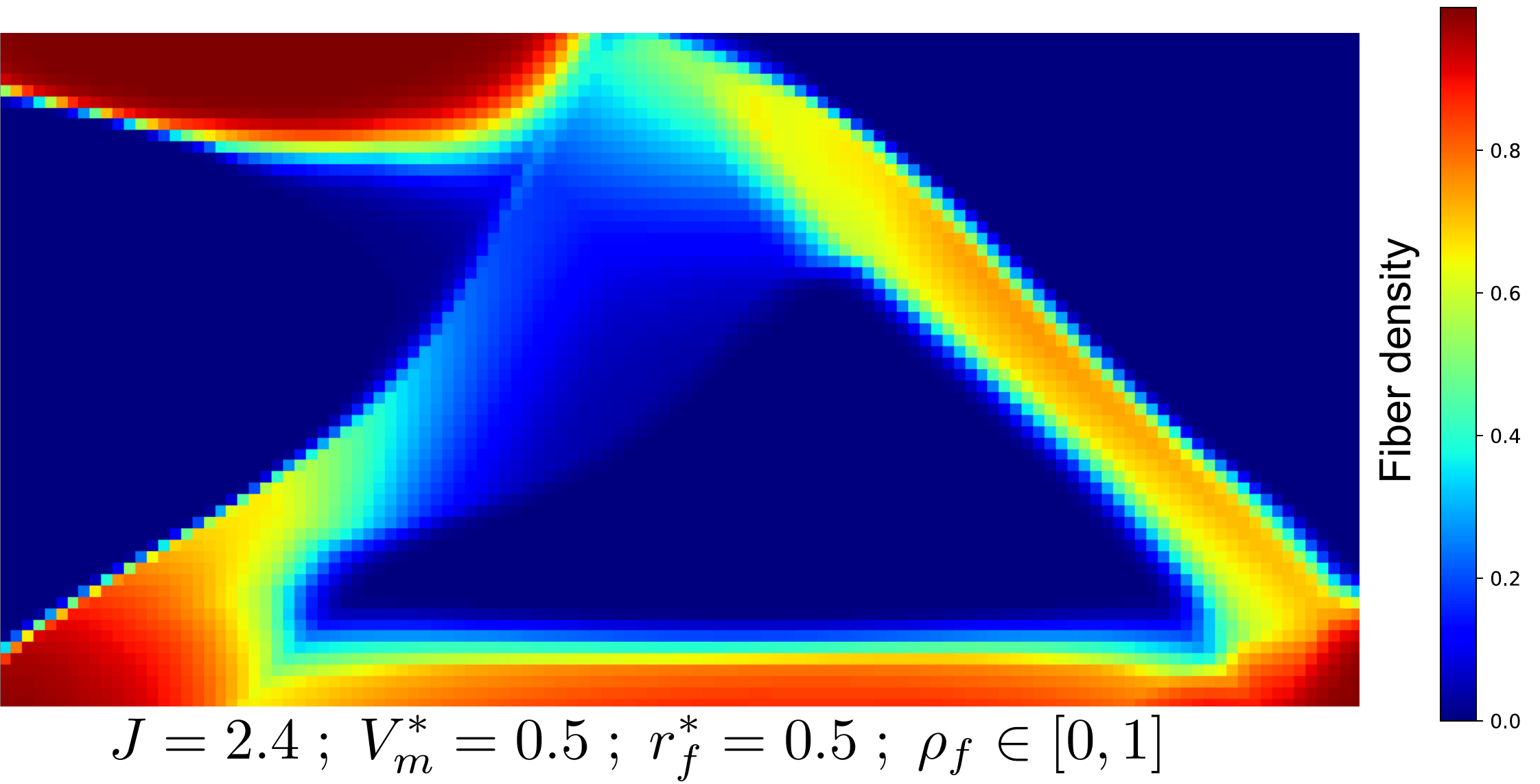}
		\caption{No bounds on fiber density}
		\label{fig:baselineNoBound}
	\end{subfigure}
	
	\begin{subfigure}[b]{\linewidth}
		\centering
		\includegraphics[width=0.5\linewidth]{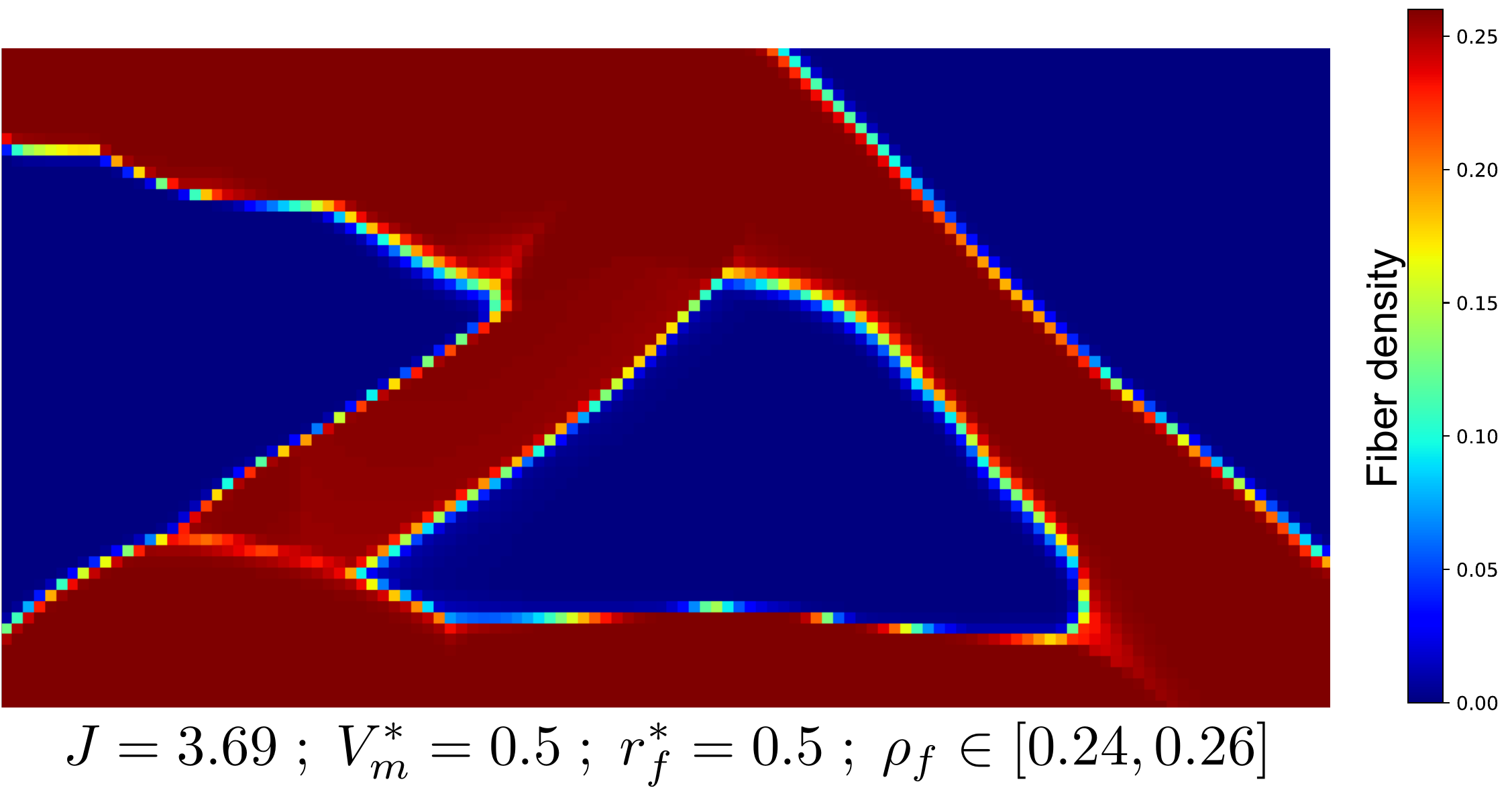}
		\caption{Constant fiber density}
		\label{fig:constantFiberDensity}
	\end{subfigure}
	
	\begin{subfigure}[b]{\linewidth}
		\centering
		\includegraphics[width=0.5\linewidth]{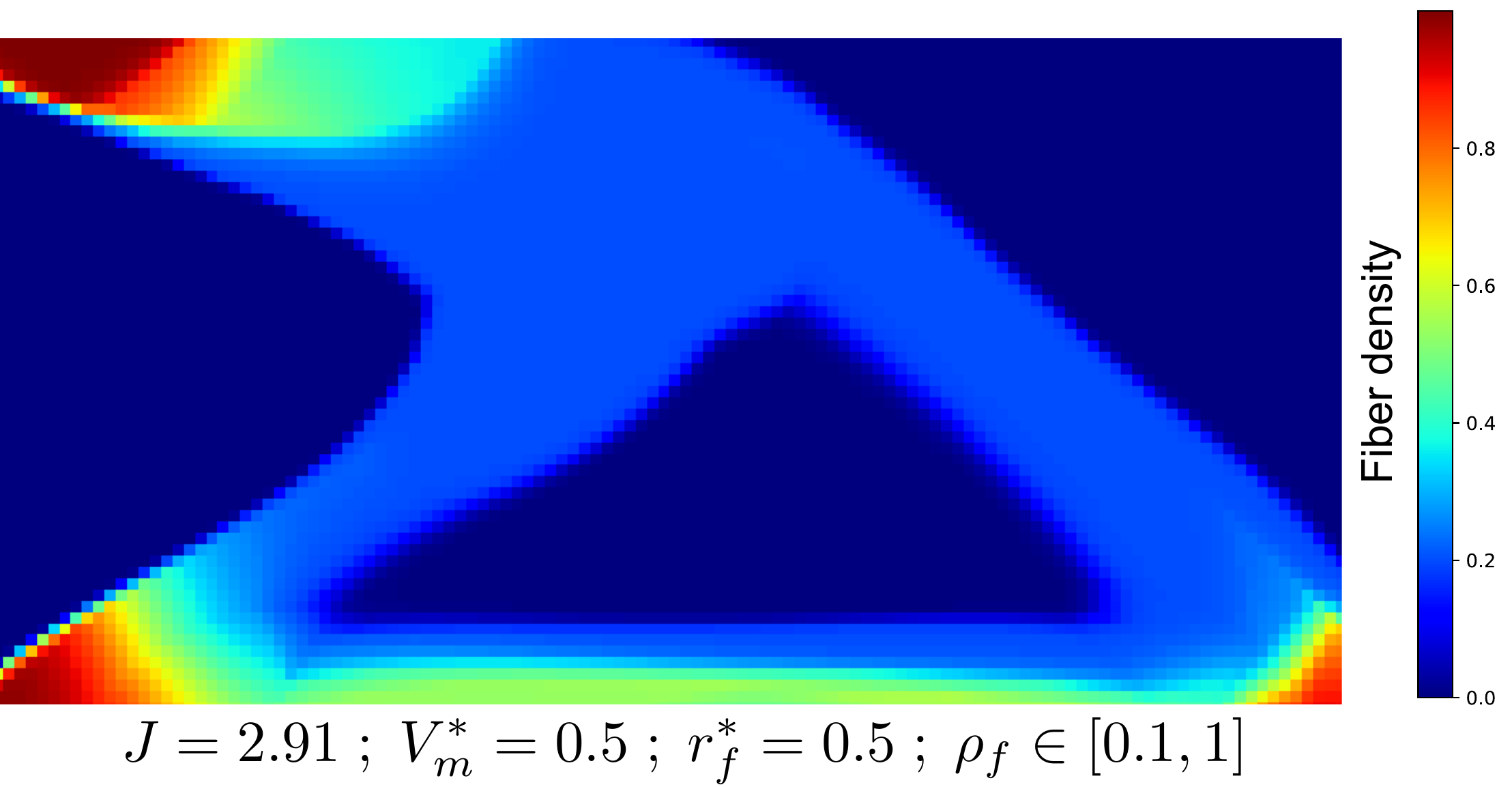}
		\caption{Lower bound on fiber density}
		\label{fig:lowerBoundFiberDensity}
	\end{subfigure}
	
	\begin{subfigure}[b]{\linewidth}
		\centering
		\includegraphics[width=0.5\linewidth]{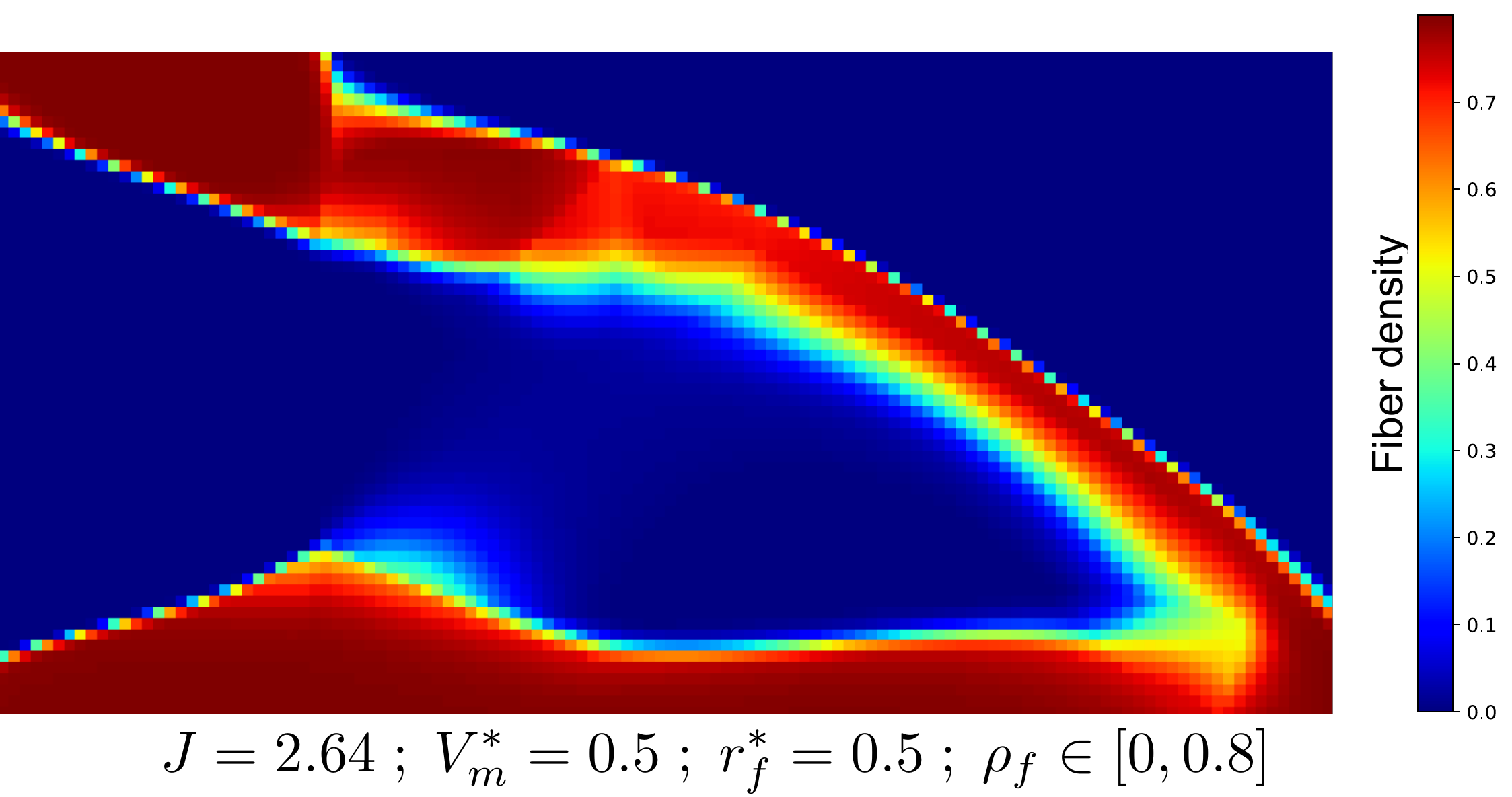}
		\caption{Upper bound on fiber density}
		\label{fig:upperBoundFiberDensity}
	\end{subfigure}
	
	\caption{Design of a tip-cantilever with manufacturing considerations}
	\label{fig:manufConsFiberDensity}
\end{figure}

 While one may obtain continuous fibers using the proposed algorithm, additional constraints may often be required to render the obtained designs manufacturable. For instance, the heavy clustering of fibers at the lower left in \Cref{fig:tipCantPrinted} may not be actually manufacturable  by continuous fiber printers. Further, the fibers may come infused/cladded with polymer matrix, making it possible to produce designs with only constant fiber density. Finally, there could be lower bounds on the fiber density owing to manufacturing limitations. Here we showcase how one may overcome these difficulties by imposing lower and upper bounds on the fiber density. In particular, the fiber density from the NN can further be transformed as $\rho_f \leftarrow \rho_f^L + (\rho_f^U - \rho_f^L) \rho_f$, where $\rho_f^L$ and $\rho_f^U$ are the lower and upper bounds on the fiber density respectively. \Cref{fig:manufConsFiberDensity} showcases the fiber densities with such imposed constraints. While we have outlined a few manufacturing constraints, other constraints may be imperative such as constraining the curvature of the fiber path, accessibility of the domain \cite{mirzendehdel2019exploring,Mirzendehdel2020AccessibilityTO,mirzendehdel2022topology}, distinction of shell and infill \cite{Wu2017ShellInfill} substructures remain to be explored.

\subsection{Computational Cost}
\label{sec:expts_cost}

\begin{figure}[!ht]
	\centering
	\begin{subfigure}[b]{\linewidth}
		\centering
		\includegraphics[width=0.5\linewidth]{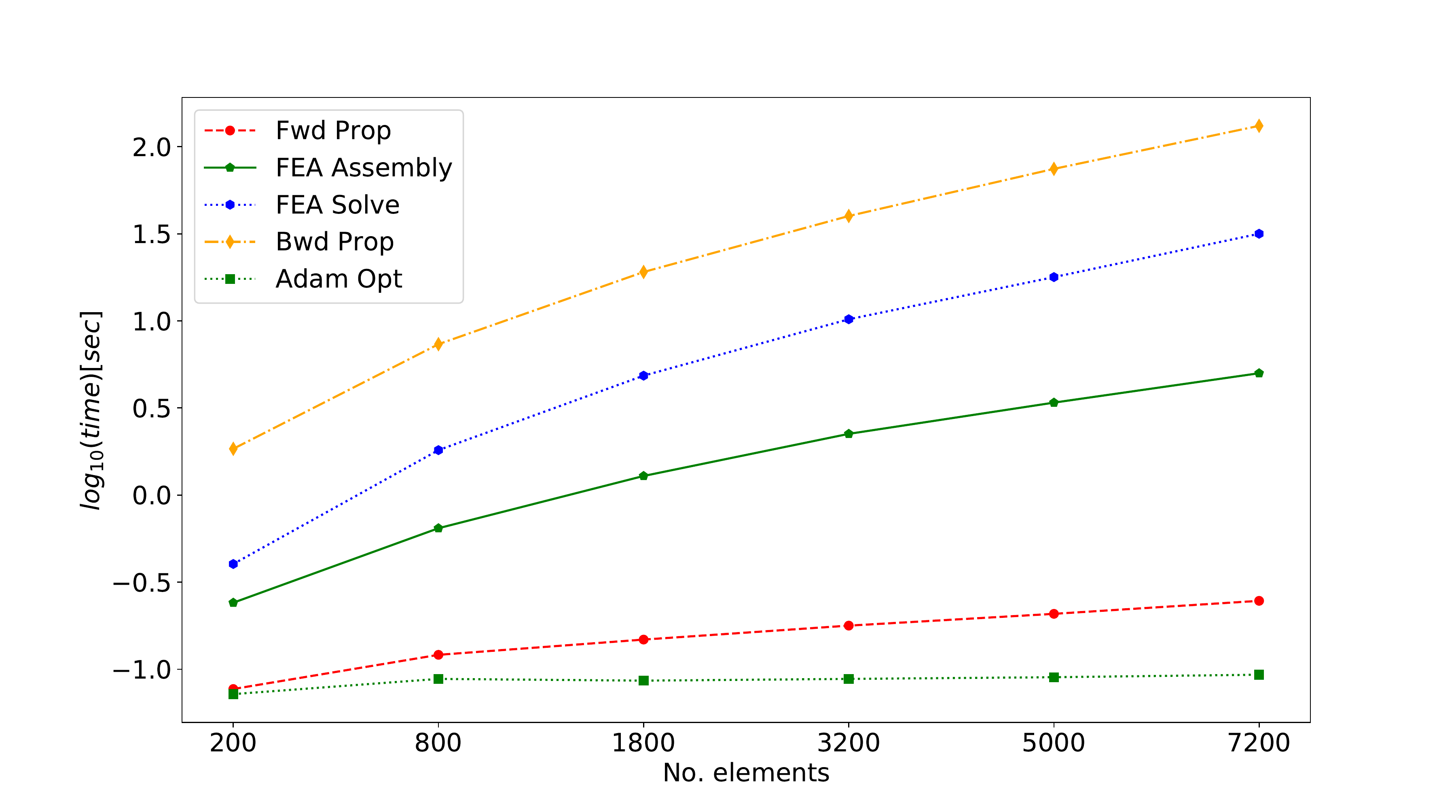}
		\caption{CPU}
		\label{fig:time_CPU}
	\end{subfigure}
	
	\begin{subfigure}[b]{\linewidth}
		\centering
		\includegraphics[width=0.5\linewidth]{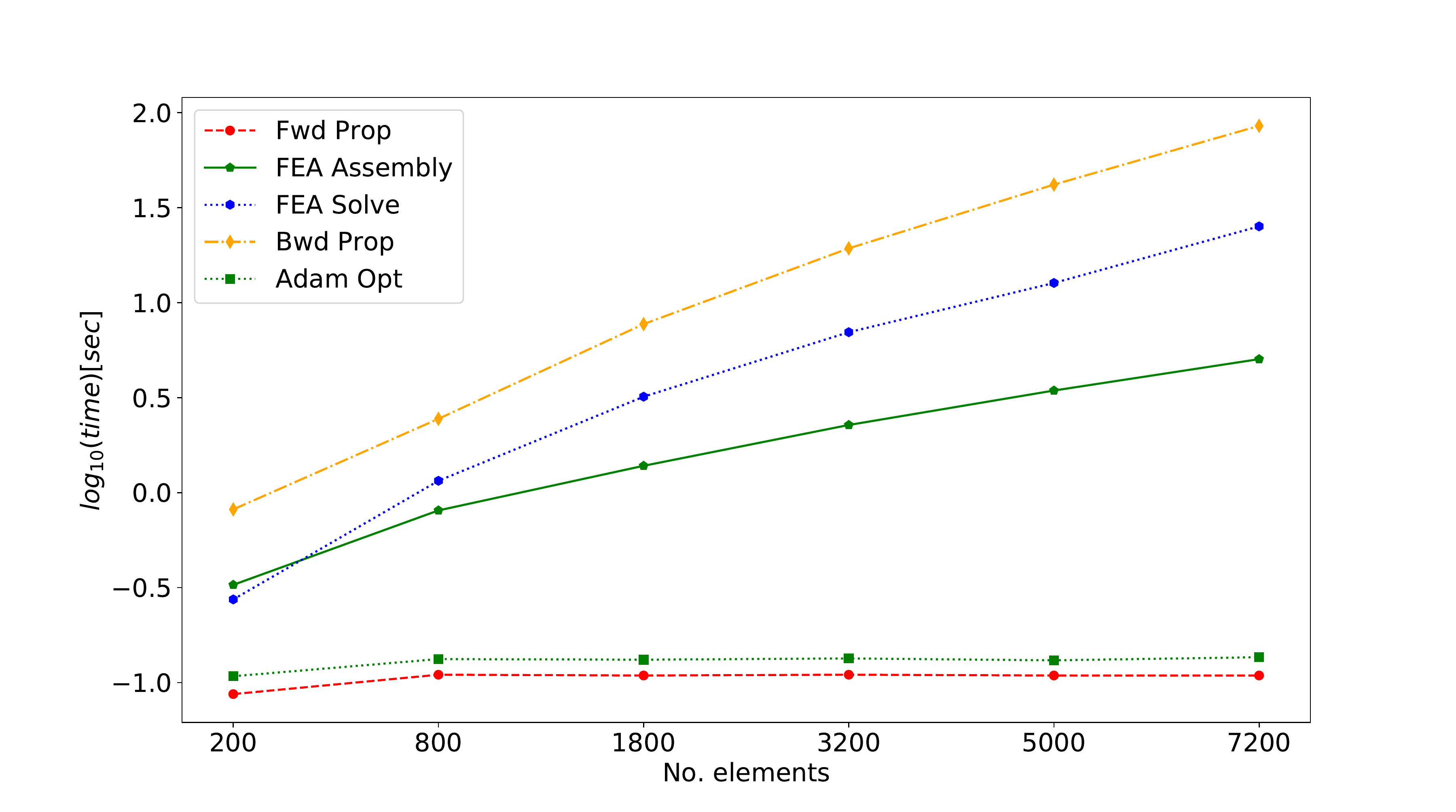}
		\caption{GPU}
		\label{fig:time_GPU}
	\end{subfigure}
	\caption{Cost of various operations for varying number of elements.}
	\label{fig:computationTime}
\end{figure}

We briefly summarize the computational costs with varying number of finite elements; all other parameters were kept constant at default values. We quantify the time taken for various operations in the optimization loop (\ref{alg:FRC-TOuNN}): namely, forward propagation (\Cref{alg:FRC-TOuNN}.\ref{alg:fwdPropNN}), stiffness matrix assembly (\Cref{alg:FRC-TOuNN}.\ref{alg:elasticityTensorCompute}, \Cref{alg:FRC-TOuNN}.\ref{alg:elemStiffnessCompute}), sparse FEA solve (\Cref{alg:FRC-TOuNN}.\ref{alg:feSolve} -  \Cref{alg:FRC-TOuNN}.\ref{alg:lossCompute}), backward propagation (\Cref{alg:FRC-TOuNN}.\ref{alg:autoDiff}), and Adam optimizer (\Cref{alg:FRC-TOuNN}.\ref{alg:adamStep}). The computations are on a desktop machine with an
\textsf{Intel}$^{\small{\textregistered}}$
\textsf{Core}$^{\small{\texttrademark}}$i7-7820X CPU with 8 processors running
at 4.5 GHz, 32 GB of host memory, and an
\textsf{NVIDIA}$^{\small{\textregistered}}$
\textsf{GeForce}$^{\small{\textregistered}}$ GTX 1080 GPU with 2,560 \textsf{CUDA} cores and 8 GB of device memory. The result summed over 200 iterations is summarized in \Cref{table:computationTime}. The time taken for individual operations on the CPU and GPU is illustrated in \Cref{fig:time_CPU} and \Cref{fig:time_GPU} respectively. We observe that the computational cost is dominated by backward-propagation. Note that the back-propagation is through the entire computation chain \Cref{fig:fwdAndBwdComputation}. This includes one more solve of the FE system for the adjoint sensitivities as well as backprop through the NN. For a comparison of time taken by baseline TOuNN and mesh-based SIMP \cite{sigmund2001Code99}, we refer the readers to \cite{ChandrasekharTOuNN2020}. 

\begin{table}[!ht]
	\centering
	\caption{Total time taken for varying mesh size.}
	\begin{tabular}{ccc}
		\hline \hline 
		{\#Elems} & {CPU Time (sec)} & {GPU Time (sec)} \\ \hline
		200              & 2.695                   & 1.701                   \\ 
		800              & 10.092                  & 4.779                   \\ 
		1800             & 25.548                  & 12.65                   \\ 
		3200             & 52.848                  & 28.943                  \\ 
		5000             & 96.251                  & 58.326                  \\ 
		7200             & 168.898                 & 115.881                 \\ 
	\end{tabular}
	
	\label{table:computationTime}
\end{table}

\subsection{Compliant Mechanism}
\label{sec:result_compliantMech}

\begin{figure}[!ht]
	\begin{subfigure}[b]{\linewidth}
		\centering
		\includegraphics[width=0.4\linewidth]{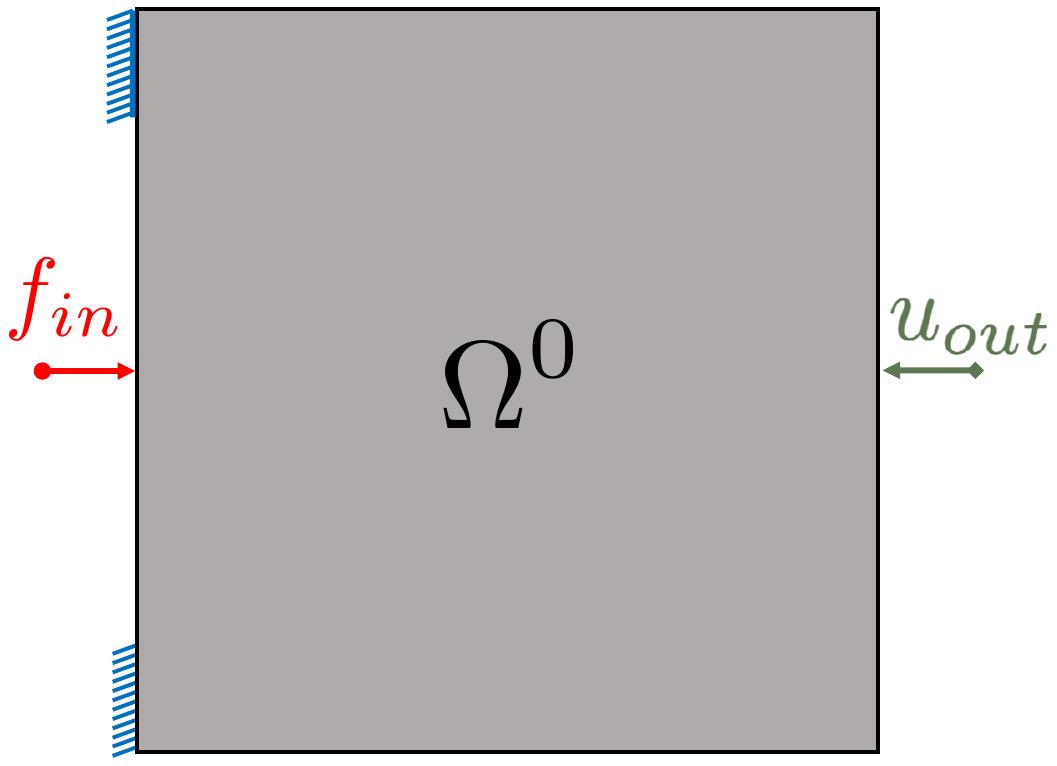}
		\caption{Loading for compliant inverter mechanism}
		\label{fig:compliantMechanismBC}
	\end{subfigure}
	\hfill 
	\begin{subfigure}[b]{\linewidth}
		\centering
		\includegraphics[width=0.3\linewidth]{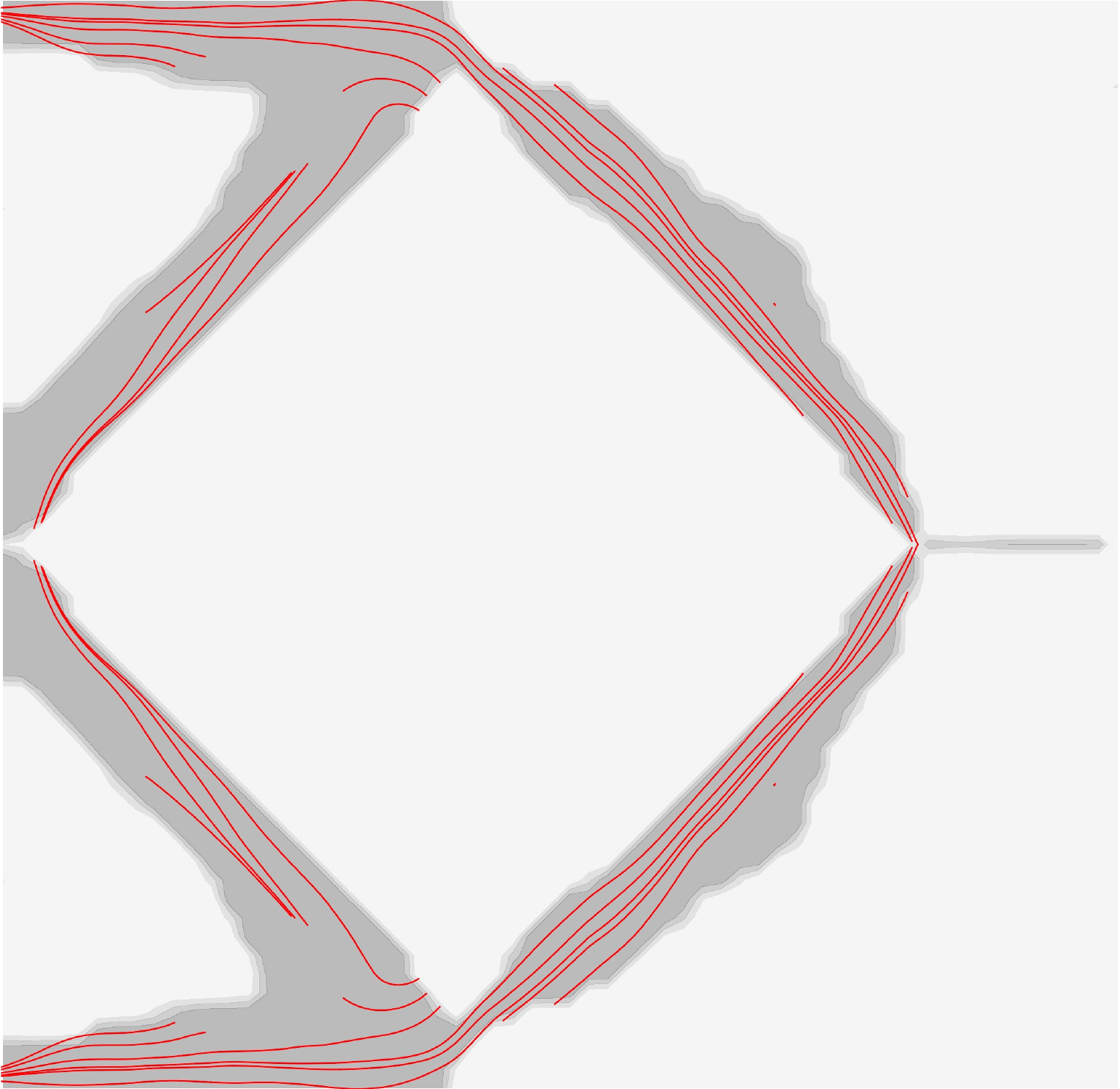}
		\caption{Optimized fiber composite compliant mechanism}
		\label{fig:compliantMechanismOptimized}
	\end{subfigure}
	\caption{Optimized fiber composite compliant inverter mechanism using FRC-TOuNN.}
\end{figure}

While TO allows us to derive designs in an automated fashion for specified requirements, a key factor during the design stages is experimenting with the optimization objectives and constraints themselves. Current trends involve error-prone derivation and laborious implementation of the objectives and their sensitivities. In this work, by expressing our computations completely using an end-to-end differentiable framework. To illustrate, we consider the optimization of compliant mechanisms. Here, given input force $f_{in}$, we are interested in maximizing the output displacement at stipulated nodes. By using automatic differentiation, one may simply change the objective in  \Cref{eq:optimization_base_objective} to \Cref{eq:compliantMechObjective}, without any other intervention
\begin{equation}
	\underset{\bm{w}}{\text{min}} \;  J = -u_{out}
	\label{eq:compliantMechObjective}
\end{equation}
The setup and the design obtained with $V_m^* = 0.3$  and $r_f^* = 0.5$ is illustrated in \Cref{fig:compliantMechanismBC} and \Cref{fig:compliantMechanismOptimized} respectively.

\subsection{Multi-Material Fibrous Composites}

\begin{figure}[!ht]
	\begin{center}
		\includegraphics[scale=0.8,trim={50 0 0 0},clip]{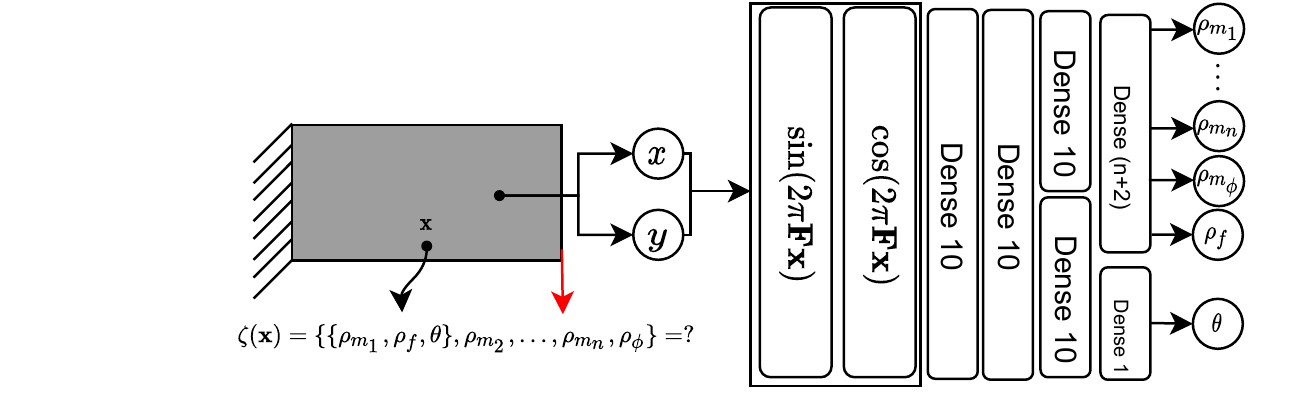}%
		\caption{Neural Network for multi-material fibrous composites. }
		\label{fig:NN_multiMatFiber}
	\end{center}
\end{figure}

FRCs comprising for different materials are also of practical interest \cite{galinska2020mechanical}. Towards this, we extend our formulation to allow for multiple materials. In particular, we suppose one of the base matrix ($m_1$) to be fiber composites and others being a homogeneous matrix. Supposing there be $n$ different allowable matrix (and with $m_{\phi}$ being void), our topological description can be expressed as $\zeta(\bm{x}) = \{ \{\rho_{m_1}, \rho_{f}, \theta \}, \rho_{m_2}, \ldots \rho_{m_n}, \rho_{m_\phi}\}$. Accordingly, the NN is modified to accommodate the new topological description (See \Cref{fig:NN_multiMatFiber}). Observe that only the output layer is modified to accommodate for the new description. The $n+2$ matrix output neurons are activated by a Softmax function to ensure physical validity of the densities via  the partition of unity (\Cref{eq:multimatFib_partitionOfUnity}) 

\begin{equation}
	\rho_{m_1} + \rho_{m_2} + \ldots + \rho_{m_n} + \rho_{m_\phi} = 1
	\label{eq:multimatFib_partitionOfUnity}
\end{equation}

\begin{equation}
	0 \leq \rho_{m_i} \leq 1 \quad i = 1,2,\ldots,n,\phi
	\label{eq:multimatFib_physicalValidityOfDensity}
\end{equation}

The  effective Elasticity tensor $[D]$ at a point $\bm{x}$  (\Cref{eq:D_matrix_net_fiber_matrix}) can then be modified as:

\begin{equation}
	\begin{aligned}
		[D(\bm{x})] =~&  \rho_{m_1}^p\bigg( \rho_f  [\mathcal{T}_1]^{-1} [D^0_f] [\mathcal{T}_2]  +  ( 1-\rho_f)[D^0_{m_0}] \bigg) \\+~& \sum\limits_{k=2}^n \rho_{m_k}^p [D^0_{m_k}]
		\label{eq:D_multiMatfiber_Dnet}
	\end{aligned}
\end{equation}
The modified optimization problem is expressed in \Cref{eq:optimization_multiMatFiber_Eqn}. We suppose the constituent materials to have a mass density $\lambda$. We then constrain the net mass of the system \Cref{eq:optimization_massConsMultiMatFiber} and the fiber volume fraction \Cref{eq:optimization_fiberVolConsMultiMatFiber}. For a detailed treatment of the multi-material formulation, we refer the readers to \cite{chandrasekhar2021multi}

\begin{subequations}
	\label{eq:optimization_multiMatFiber_Eqn}
	\begin{align}
		& \underset{\bm{\zeta}}{\text{minimize}}
		& &J(\bm{\zeta}) \label{eq:optimization_multiMatFiber_objective}\\
		& \text{subject to}
		& & \bm{K}(\bm{\zeta})\bm{u} = \bm{f}\label{eq:optimization_multiMatFiber_govnEq}\\
		& & & g_m (\bm{\rho_M})  \coloneqq \frac{\sum\limits_{k=1}^n\sum\limits_e \lambda_k \rho_{m_k}(\bm{x}_e) v_e}{m^*} - 1 \leq 0  \label{eq:optimization_massConsMultiMatFiber}\\
		& & & g_f (\bm{\rho_f}) \coloneqq \frac{\sum\limits_e \rho_f(\bm{x}_e) v_e}{ V_f^*\sum\limits_e v_e} - 1  \leq 0 \label{eq:optimization_fiberVolConsMultiMatFiber}
	\end{align}
\end{subequations}

\begin{figure}[!ht]
	\begin{subfigure}[b]{\linewidth}
		\centering
		\includegraphics[width=0.5\linewidth]{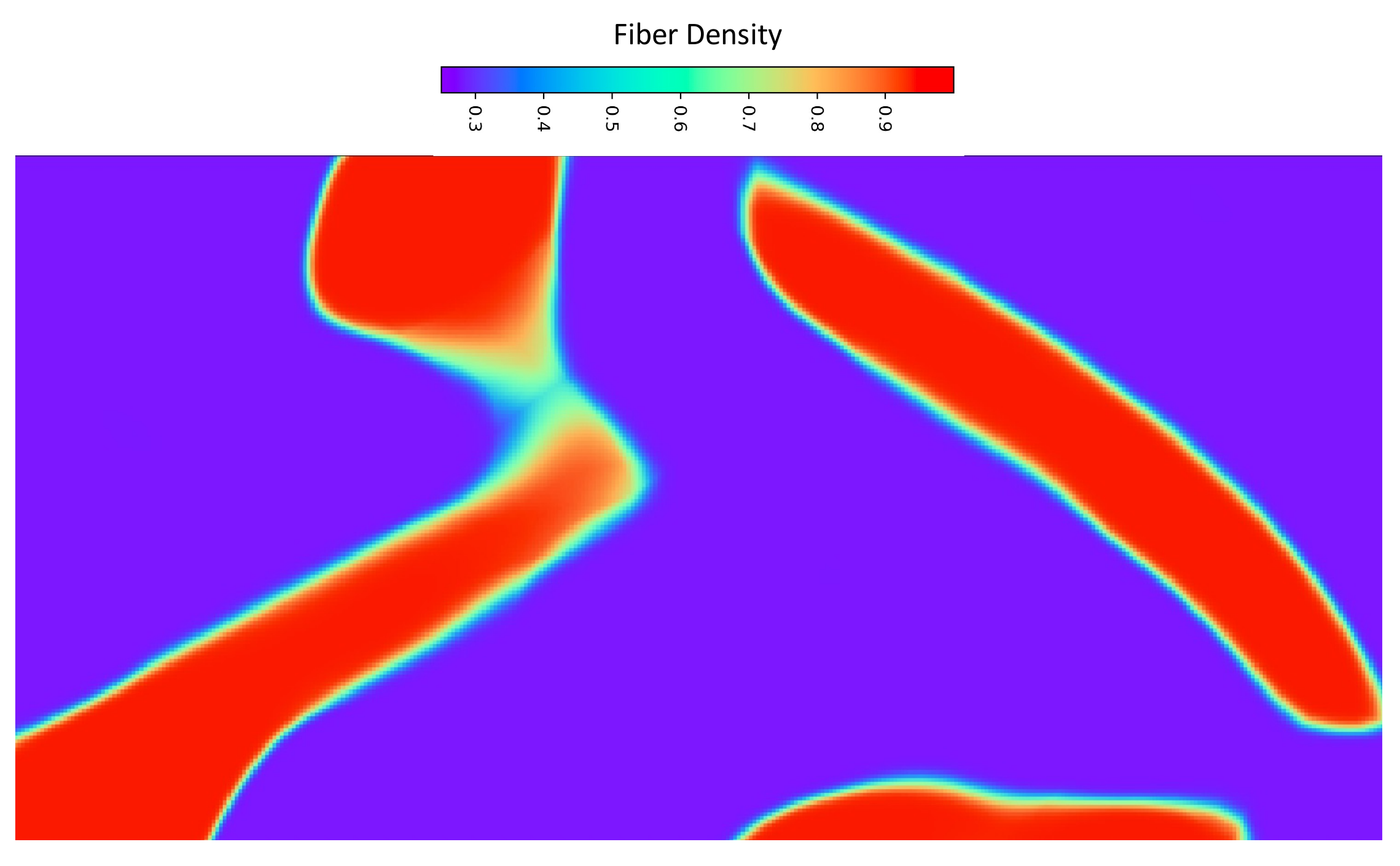}
		\caption{Fiber density}
		\label{fig:multimaterialFiberDensity}
	\end{subfigure}
	
	\begin{subfigure}[b]{\linewidth}
		\centering
		\includegraphics[width=0.5\linewidth]{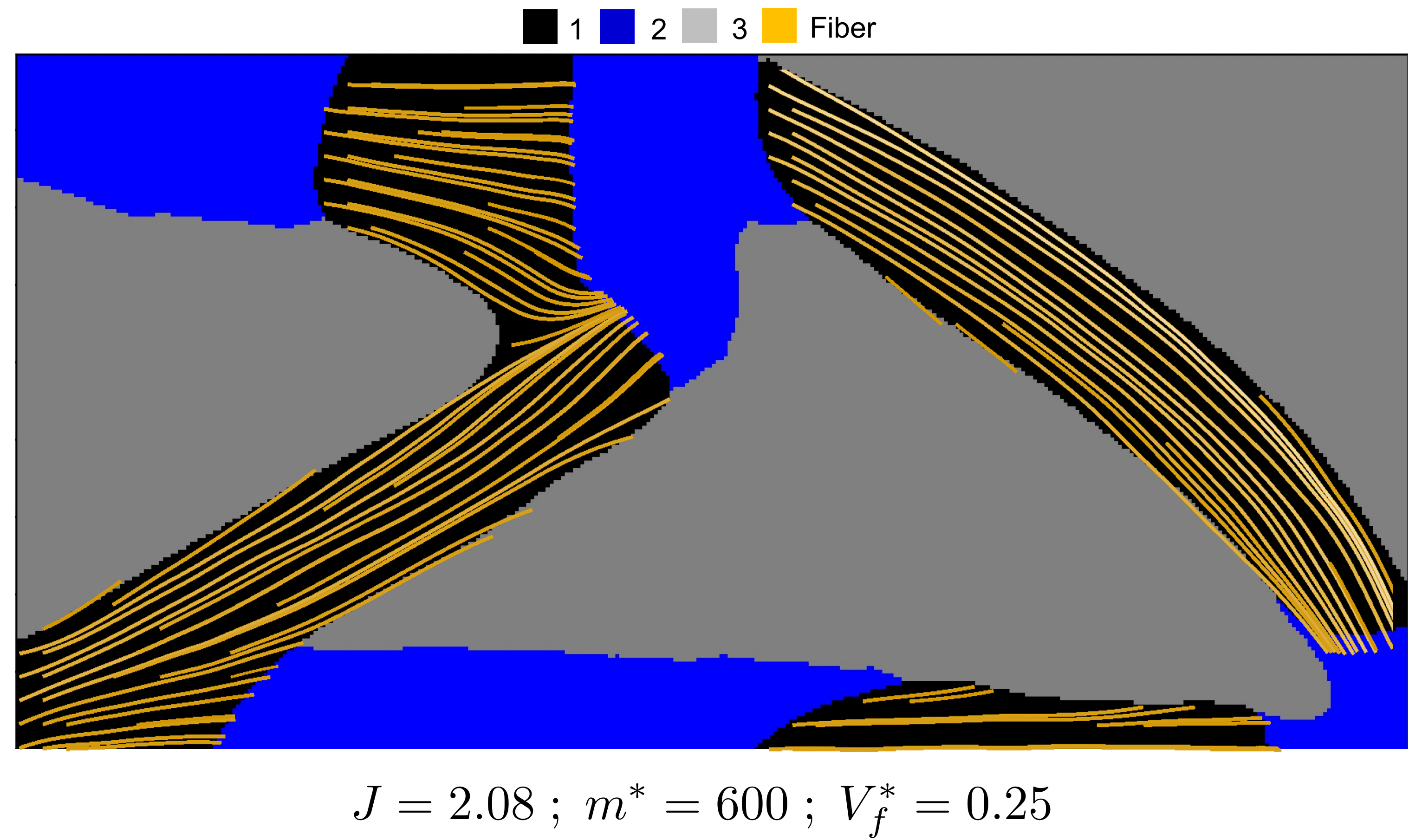}
		\caption{Multi-Material Tip Cantilever beam with fiber composites.}
		\label{fig:multiMatFiberTipCant}
	\end{subfigure}
	\caption{Design of a tip-cantilever multi-material fibrous composite}
	\label{fig:multiMaterialFibrousComposite}
\end{figure}

\begin{table}[!ht]
	\caption{Material properties for multi-material design}
	\begin{center}
		\begin{tabular}{  C{14mm}  C{24mm}  C{11mm} C{11mm} }
			\hline\hline
			Material & Color Code & $E~(Pa)$ & $\lambda~(kg/m^3)$ \\ \hline
			1 & Black (matrix) &0.6 & 0.4 \\ 
			2 &Blue & 4 & 1 \\ 
			3 & Gray (void) & 1e-9 & 1e-9 \\ 
		\end{tabular}
	\end{center}
	\label{table:multiMatProp}
\end{table}
As an example, we consider the optimization of the tip-cantilever beam. We suppose that the domain can accommodate  the  materials in \Cref{table:multiMatProp}. The base matrix 1 may be impregnated with a fiber with $E_{\parallel} = 4$, $E_{\perp} = 1$, $\nu = 0.3$ and $G = 0.7$. With a  net allowed mass of $m^* = 600$ and fiber volume fraction $V_f^* = 0.25$, we obtain the optimized topology as shown in \Cref{fig:multiMaterialFibrousComposite}.

\section{Conclusion}
\label{sec:conclusion}

The main contribution of this paper is a novel neural-network based topology optimization framework for optimizing continuous fiber reinforced composites. Representing the matrix density, fiber density and orientation implicitly via a NN, we showcased that one could extract long continuous fibers with sub-fiber spacing that reflects the fiber density. The method was validated by comparing it against published research, and characterized through several problems in 2D. The obtained designs were exported and printed on an \textit{Ultimaker 3} printer. Further, implementing  an end-to-end differentiable framework allowed for faster experimentation. This was demonstrated by applying our framework to the design of compliance minimized design, design of compliant mechanism and multi-material FRCs.  The extension of the proposed method to 3D to accommodate volumetric profiles \cite{Elber2018VolumetricPrintPaths}, multi-axis prints \cite{dai2018multiAxisPrinting} or build orientation \cite{Chandrasekhar2020Fiber} remains to be explored. Structurally validating the obtained results remains to be seen in the future \cite{rankouhi2016failure}. While long continuous fibers were obtained, no explicit measure constraining the length, continuity or curvature was imposed. Enforcing such constraints could be of practical importance \cite{shafighfard2019FiberCurvature}. We observed that back-propagation consumes a large portion of the computational cost. Exploring techniques reducing this cost would be critical for wider adoption of the proposed method. While we used a direct sparse solver for FE simulation, large scale analysis would inevitably require iterative solver. Either integrating such solvers in automatic differentiation framework or online learning of FE solutions \cite{chi2021universal} and data-driven FE approaches \cite{rade2021algorithmically} \cite{banga20183d} remains to be explored. Future work will also focus on an improved post-processing algorithm.

\section*{Compliance with ethical standards}
The authors declare that they have no conflict of interest.

\bibliographystyle{plainnat}
\bibliography{references}  

\end{document}